\documentclass[letterpaper]{JHEP3}

\usepackage{amsfonts}
\usepackage{amssymb}
\usepackage{epsfig}
\usepackage{graphicx}
\usepackage{amsmath}
\usepackage{amsxtra}
\usepackage{color}

\def\circa#1{\,\raise.3ex\hbox{$#1$\kern-.75em\lower1ex\hbox{$\sim$}}\,}
\makeatletter
\def\art{\@ifnextchar[{\eart}{\oart}}
\def\eart[#1]#2#3#4#5#6{{\rm #2}, {\em #3  #4} {\rm (#6) #5} ({\em #1})}
\def\hepart[#1]#2{{\rm #2, \em#1}}
\newcommand{\oart}[5]{{\rm #1}, {\em #2  #3} {\rm (#5) #4}}

\newcounter{alphaequation}[equation]
\def\thealphaequation{\theequation\hbox to
0.6em{\hfil\alph{alphaequation}\hfil}}
\def\eqnsystem#1{
\def\@eqnnum{{\rm (\thealphaequation)}}
\def\@@eqncr{\let\@tempa\relax \ifcase\@eqcnt \def\@tempa{& & &} \or
\def\@tempa{& &}\or \def\@tempa{&}\fi\@tempa
\if@eqnsw\@eqnnum\refstepcounter{alphaequation}\fi
\global\@eqnswtrue\global\@eqcnt=0\cr}
\refstepcounter{equation} \let\@currentlabel\theequation \def\@tempb{#1}
\ifx\@tempb\empty\else\label{#1}\fi
\refstepcounter{alphaequation}
\let\@currentlabel\thealphaequation
\global\@eqnswtrue\global\@eqcnt=0 \tabskip\@centering\let\\=\@eqncr
$$\halign to \displaywidth\bgroup \@eqnsel\hskip\@centering
$\displaystyle\tabskip\z@{##}$&\global\@eqcnt\@ne
\hskip2\arraycolsep\hfil${##}$\hfil& \global\@eqcnt\tw@\hskip2\arraycolsep
$\displaystyle\tabskip\z@{##}$\hfil
\tabskip\@centering&\llap{##}\tabskip\z@\cr}
\def\endeqnsystem{\@@eqncr\egroup$$\global\@ignoretrue} \makeatother

\def\be{\begin{equation}}
\def\ee{\end{equation}}
\def\bea{\begin{eqnarray}}
\def\eea{\end{eqnarray}}

\newcommand{\vo}{{\cal V}}

\newcommand{\roughly}[1]{\mathrel{\raise.3ex\hbox{$#1$\kern-0.85em
\lower1ex\hbox{$\sim$}}}}

\def\ba{\begin{eqnarray}}
\def\ea{\end{eqnarray}}
\def\be{\begin{equation}}
\def\ee{\end{equation}}

\def\KK{{\scriptscriptstyle KK}}

\def\10{{\scriptscriptstyle 10}}
\def\4{{\scriptscriptstyle 4}}
\def\GUT{{\scriptscriptstyle GUT}}

\def\nn{\nonumber}
\def\mc{\mathcal}

\def\({\left(}
\def\){\right)}

\title{Heterotic Moduli Stabilisation}

\author{
M.~Cicoli,${}^{1,2,3}$
S.~de Alwis,${}^{3,4}$  and A.~Westphal${}^5$\\

$^1$ Dipartimento di Fisica ed Astronomia, Universit\`a di Bologna, Bologna, Italy. \\
$^2$ INFN, Sezione di Bologna, Italy.\\
$^3$ Adbus Salam ICTP, Strada Costiera 11, Trieste 34014, Italy.\\
$^4$ UCB 390 Physics Dept. University of Colorado, Boulder CO 80309, U.S.A.\\
$^5$ Deutsches Elektronen-Synchrotron DESY, Theory Group, D-22603 Hamburg, Germany}

\abstract{We perform a systematic analysis of moduli stabilisation for weakly coupled heterotic string theory compactified
on internal manifolds which are smooth Calabi-Yau three-folds up to $\alpha'$ effects.
We first review how to stabilise all the geometric and gauge bundle moduli
in a supersymmetric way by including fractional fluxes, the requirement of a holomorphic gauge bundle,
D-terms, higher order perturbative contributions to the superpotential
as well as non-perturbative and threshold effects. We then show that the inclusion of $\alpha'$
corrections to the K\"ahler potential leads to new stable Minkowski (or de Sitter) vacua
where the complex structure moduli and the dilaton are fixed
supersymmetrically at leading order, while the stabilisation of the K\"ahler moduli
at a lower scale leads to spontaneous breaking supersymmetry.
The minimum lies at moderately large volumes of all the geometric moduli, at perturbative values of the string coupling
and at the right phenomenological value of the GUT gauge coupling.
We also provide a dynamical derivation of anisotropic compactifications with stabilised moduli
which allow for perturbative gauge coupling unification around $10^{16}\,{\rm GeV}$.
The value of the gravitino mass can be anywhere between the GUT and the TeV scale
depending on the stabilisation of the complex structure moduli.
In general, these are fixed by turning on background fluxes,
leading to a gravitino mass around the GUT scale since
the heterotic three-form flux does not contain enough freedom to tune the superpotential to small values.
Moreover accommodating the observed value of the cosmological constant is a challenge.
Low-energy supersymmetry could instead be obtained
by focusing on particular Calabi-Yau constructions where the gauge bundle is holomorphic only
at a point-like sub-locus of complex structure moduli space,
or situations with a small number of complex structure moduli
(like orbifold models), since in these cases one may fix all the moduli
without turning on any quantised background flux.
However obtaining the right value of the cosmological constant is even more of a challenge in these cases.
Another option would be to focus on compactifications on non-complex manifolds,
since these allow for new geometric fluxes which could be used to tune
the superpotential as well as the cosmological constant, even if the moduli space of these manifolds is presently only poorly understood.}

\preprint{DESY--13-066}

\begin{document}

\section{Introduction}
\label{Introduction}

String theory is a candidate for a quantum theory of gravity with full unification of the forces of nature. As such it should be able to describe the patterns of the Standard Models (SMs) of particle physics and cosmology. For this description of 4D physics, string theory needs to compactify its ambient 10D space-time. The multitude of possible compactification choices together with a plethora of massless 4D `moduli' fields originating from the deformation modes of the extra dimensions, leads to vacuum degeneracy and moduli problems. Recent progress in achieving moduli stabilisation points to the possibility of an exponentially large set of cosmologically distinct de Sitter (dS) solutions of string theory with positive but tiny cosmological constant, the `landscape' (for reviews see \cite{Grana:2005jc,Douglas:2006es}).

These results need to be combined with string constructions of viable particle physics. One fruitful region of the string landscape for this purpose is weakly coupled heterotic string theory. Recent works on heterotic compactifications on both smooth Calabi-Yau (CY) manifolds~\cite{Braun:2005ux} and their singular limits in moduli space, orbifolds~\cite{Buchmuller:2005jr,Buchmuller:2006ik,Lebedev:2006kn,Lebedev:2007hv,Lebedev:2008un}, provided constructions of 4D low-energy effective field theories matching the minimal supersymmetric version of the SM (MSSM) almost perfectly. However, in contrast to the understanding achieved in type IIB string theory, heterotic CY or orbifold compactifications lack a well controlled description of moduli stabilisation, and consequently, of inflationary cosmology as well.\footnote{However, for some recent attempts see e.g.~\cite{Dundee:2010sb,Parameswaran:2010ec}.}

As weakly coupled heterotic CY compactifications lack both D-branes and a part of the three-form flux available in type IIB, historically moduli stabilisation in the heterotic context focused mostly on the moduli dependence of 4D non-perturbative contributions to the effective action from gaugino condensation~\cite{Font:1990nt,Ferrara:1990ei,Nilles:1990jv}. While this produced models of partial stabilisation of the dilaton
and some K\"ahler moduli~\cite{Dundee:2010sb,Casas:1990qi,de Carlos:1992da,Gallego:2008sv}, this route generically failed at describing controlled and explicit stabilisation of the ${\cal O}(100)$ complex structure moduli of a given CY.
Moreover, the resulting vacua tend to yield values of the compactification radius and string coupling
(given by the dilaton) at the boundary of validity of the supergravity approximation and the weak coupling regime.

The works \cite{GKLM,BdA,CKL} proposed to include the three-form flux $H$ to stabilise the complex structure moduli
in combination with hidden sector gaugino condensation for supersymmetric dilaton stabilisation.
The inclusion of fluxes in the heterotic string was originally studied by Strominger \cite{Strominger:1986uh}
who showed that, by demanding $\mc{N}=1$ supersymmetry, the classical 10D equations of motion imply
$H=-\frac{i}{2}(\partial-\bar{\partial})J$ where $J$ is the fundamental $(1,1)$-form on the internal space.
Hence a non-vanishing three-form flux breaks the K\"ahler condition $dJ=0$. Note that this is the
case of $(0,2)$-compactifications which allow for MSSM-like model building and the generation of worldsheet instantons,
since in the non-standard embedding the Chern-Simons term gives a non-zero contribution to the three-form flux $H$.
However this contribution is at order $\alpha'$, implying that the Calabi-Yau condition is preserved at tree-level
and broken only at order $\alpha'$. Moreover, in the heterotic case, due to the absence of Ramond-Ramond three-form fluxes,
there is generically no freedom to tune the superpotential small enough to fix the dilaton at weak coupling.
However, a sufficiently small superpotential could be obtained by considering fractional Chern-Simons invariants
(such as discrete Wilson lines) \cite{GKLM}. Note that it is natural to take these effects into account
for $(0,2)$-compactifications which are the most relevant for both model building and moduli stabilisation
since, as we pointed out above, they feature a non-vanishing Chern-Simons contribution to $H$, regardless of the presence of fractional Chern-Simons invariants.\footnote{As we shall describe in section \ref{Sec311}, the co-exact piece
of the Chern-Simons term is responsible for the breaking of the K\"ahler condition $dJ=0$ while the generation of fractional invariants
is controlled by the harmonic piece of the Chern-Simons term.}

Supersymmetric vacua with all geometric moduli stabilised could be achieved by fixing the K\"ahler moduli
via contributions from threshold corrections to the gauge kinetic function~\cite{Dixon:1990pc,BHW}.
However this minimum cannot be trusted since it resides in a strong coupling regime where the gauge coupling is even driven into negative values \cite{GKLM}.
The inclusion of a single worldsheet instanton contribution can resolve this difficulty \cite{CKL}.
However, none of these vacua break supersymmetry, resulting in unrealistic anti-de Sitter (AdS) solutions.

In this paper, we shall present new stable Minkowski (or de Sitter) vacua where all geometric
moduli are stabilised and supersymmetry is broken spontaneously along the K\"ahler moduli directions.
Let us summarise our main results:
\begin{itemize}
\item We identify two small parameters, one loop-suppressed and the other volume-suppressed, which allow us
to expand the scalar potential in a leading and a subleading piece. This separation of scales
allows us to perform moduli stabilisation in two steps.

\item The leading scalar potential is generated by D-terms, quantised background fluxes (if needed for the stabilisation of the
complex structure moduli), perturbative contributions to the superpotential and gaugino condensation.
This potential depends on the gauge bundle moduli, the complex structure moduli and the dilaton
which are all fixed supersymmetrically at leading order.

\item The subleading scalar potential depends on the K\"ahler moduli and it is generated
by threshold corrections to the gauge kinetic function, worldsheet instantons and
$\mc{O}(\alpha'^2)$ \cite{AQS}, and $\mc{O}(\alpha'^3)$ \cite{StandardAlphaPrime,BBHL} corrections to the K\"ahler potential.
These effects give rise to new Minkowski vacua (assuming the fine-tuning problem can be solved) which break supersymmetry
spontaneously along the K\"ahler moduli directions.
The dilaton is stabilised at a value $\rm{Re}(S)\simeq 2$ in a way compatible with gauge coupling unification,
while the compactification volume is fixed at $\vo\simeq 20$ which is the upper limit compatible with string perturbativity.
These new minima represent a heterotic version of the type IIB LARGE Volume Scenario (LVS) \cite{LVS,GeneralLVS}.

\item By focusing on CY manifolds with K3- or $T^4$-fibres over a $\mathbb{P}^1$ base, we shall also show that this
LVS-like moduli stabilisation mechanism allows for anisotropic constructions where the overall volume
is controlled by two larger extra dimensions while the remaining four extra dimensions remain smaller.
This anisotropic setup is particularly interesting phenomenologically,
as it allows one to match the effective string scale to the GUT scale of gauge coupling unification~\cite{Hebecker:2004ce,Dundee:2008ts},
and fits very well with the picture of intermediate 6D orbifold GUTs emerging from heterotic MSSM orbifolds~\cite{Dundee:2008ts,Buchmuller:2007qf}.

\item The soft terms generated by gravity mediation feature universal scalar masses, A-terms and $\mu/B\mu$-term of
order the gravitino mass, $m_{3/2} = |W| M_P/\sqrt{2\rm{Re}(S)\vo}$, and suppressed gaugino masses at the \%-level.
In turn, the value of the supersymmetry breaking scale depends on the stabilisation of the complex structure moduli:
\begin{enumerate}
\item If the complex structure moduli are fixed by turning on quantised background fluxes,
due to the lack of tuning freedom in the heterotic three-form flux,
$|W|$ can at most be made of order $|W|\simeq \mc{O}(0.1-0.01)$ by turning on only Chern-Simons fractional fluxes.
Hence the gravitino mass becomes of order $M_\GUT\simeq 10^{16}$ GeV for $\rm{Re}(S)\simeq 2$
and $\vo\simeq 20$, leading to high scale supersymmetry breaking.

\item If the complex structure moduli are fixed without turning on quantised background fluxes,
the main contribution to $|W|$ can come from higher order perturbative operators or gaugino condensation.
Hence $|W|$ can acquire an exponentially small value, leading to low-scale supersymmetry \cite{HPN,Kappl:2008ie}.
\end{enumerate}
\end{itemize}
Let us discuss the stabilisation of the complex structure moduli in more detail.
In a series of recent papers \cite{AGLO1,AGLO2,Anderson:2011ty}, it has been shown that
in particular examples one could be able to fix all the complex structure moduli without the need to turn on
any quantised background flux. Note that, as we explained above, if one focuses on $(0,2)$-compactifications,
this observation is not important for preserving the CY condition (since this is broken at order $\alpha'$ regardless
of the presence of a harmonic quantised flux) but it is instead crucial to understand
the order of magnitude of the superpotential which sets the gravitino mass scale once supersymmetry is broken.
Following the original observation of Witten \cite{Witten:1985bz}, the authors of \cite{AGLO1,AGLO2,Anderson:2011ty} proved
that, once the gauge bundle is required to satisfy the Hermitian Yang-Mills equations,
the combined space of gauge bundle and complex structure moduli is not a simple direct product
but acquires a `cross-structure'. Denoting the gauge bundle moduli as $C_i$, $i=1,...,N$,
and the complex structure moduli as $Z_\alpha$, $\alpha=1,...,h^{1,2}$, this observation implies
that the dimensionality of the gauge bundle moduli space is actually a function of the complex
structure moduli, i.e. $N=N(Z)$, and viceversa the number of massless $Z$-fields actually depends
on the value of the gauge bundle moduli. As a simple intuitive example, consider a case with just one gauge bundle modulus
and a leading order scalar potential which looks like:
\be
V = \left(\sum_{\beta=1}^{h^{1,2}_{\rm fix}} |Z_\beta|^2\right) |C|^2\,.
\ee
The form of this potential implies that:
\begin{itemize}
\item If $C$ is fixed by some stabilisation mechanism (like D-terms combined with higher order $C$-dependent
terms in the superpotential) at $\langle C\rangle\neq 0$,
then $h^{1,2}_{\rm fix}$ complex structure moduli are fixed at $\langle Z_\beta\rangle = 0$
$\forall\,\beta=1,...,h^{1,2}_{\rm fix}$. Hence the number of $Z$-moduli left flat is given by
$h^{1,2}_{\rm hol}=h^{1,2}-h^{1,2}_{\rm fix}$, which is also the dimensionality of the sub-locus in complex structure
moduli space for $C\neq0$ where the gauge bundle is holomorphic. Hence the best case scenario is when this sub-locus
is just a point, i.e. $h^{1,2}_{\rm hol}=0$.

\item If the $Z$-moduli are fixed by some stabilisation mechanism (like by turning on background quantised fluxes) at
values different from zero, then the gauge bundle modulus $C$ is fixed at $\langle C\rangle=0$.\footnote{See also \cite{Donagi} 
for a mathematical discussion of this issue which basically comes to the same conclusion that gauge bundle moduli are generically absent.}
\end{itemize}
However this stabilisation mechanism generically does not lead
to the fixing of \textit{all} complex structure moduli due to the difficulty of finding examples
with $h^{1,2}_{\rm hol}=0$, i.e. with a point-like sub-locus in complex structure moduli space
where the gauge bundle is holomorphic. In fact, there is so far no explicit example in the literature
where $h^{1,2}_{\rm hol}=0$ can be obtained without having a singular CY even if there has been
recently some progress in understanding how to resolve these singular point-like sub-loci \cite{AGLO3}.
Moreover, let us stress that even if one finds a non-singular CY example with $h^{1,2}_{\rm hol}=0$
(there is in principle no obstruction to the existence of this best case scenario),
all the complex structure moduli are fixed only if $C\neq 0$, since for $C=0$ the $Z$-directions
would still be flat. As we pointed out above, $C\neq 0$ could be guaranteed by the interplay of D-terms and
higher order terms in the superpotential, but in the case when the number of $C$-moduli is large, one
should carefully check that all of them are fixed at non-zero values (for example, one might like to have some
of them to be fixed at zero in order to preserve some symmetries relevant for phenomenology like $U(1)_{B-L}$).
Thus the requirement of a holomorphic gauge bundle generically fixes some complex structure moduli but not all of them.
Note also that these solutions are not guaranteed to survive for a non-vanishing superpotential,
since one would then need to solve a set of non-holomorphic equations.

Let us therefore analyse the general case where some $Z$-moduli are left flat after the requirement of a holomorphic gauge bundle,
and summarise our results for their stabilisation:
\begin{itemize}
\item Given that promising phenomenological model building requires us to focus on the non-standard embedding where
the $H$-flux already gets a non-vanishing contribution from the co-exact piece of the Chern-Simons term,
we consider quite natural the option to turn on also a harmonic Chern-Simons piece that could yield a fractional
$Z$-dependent superpotential that lifts the remaining complex structure moduli \cite{GKLM}.

\item If $H\neq 0$, as in the case of $(0,2)$-compactifications,
both the dilaton and the warp factor could depend on the internal coordinates.
For simplicity, we shall however restrict to the solutions where both of them are constant,
corresponding to the case of `special Hermitian manifolds' \cite{Lopes Cardoso:2002hd}.

\item The inclusion of quantised background fluxes cannot fix the remaining $h^{1,2}_{\rm hol}>0$
complex structure moduli in a supersymmetric way with, at the same time, a vanishing flux superpotential $W_0$.
In fact, setting the F-terms of the $Z$-moduli to zero corresponds to setting the $(1,2)$-component of $H$ to zero,
whereas setting $W_0=0$ implies a vanishing $(3,0)$-component of $H$. As a consequence, given that the flux is real,
the entire harmonic flux $H$ is zero, and so the $h^{1,2}_{\rm hol}>0$ $Z$-moduli are still flat.\footnote{This statement is
also implicit in \cite{BdA}.} Note that this would not be the case in type IIB where the three-form flux is complex
(because of the presence of also Ramond-Ramond fluxes) \cite{GKP}.

\item The remaining $h^{1,2}_{\rm hol}>0$ $Z$-moduli can be fixed only if $W_0\neq 0$
but this would lead to a runaway for the dilaton if $W_0$ is not fine-tuned to exponentially small values
to balance the dilaton-dependent contribution from gaugino condensation.
However, due to the absence of Ramond-Ramond fluxes, the heterotic $H$-flux
does not contain enough freedom to tune $W_0$ to small values, since it is used mostly
to stabilise the complex structure moduli in a controlled vacuum.
There are then two options:
\begin{enumerate}
\item Models with either accidentally cancelling integer flux quanta or only Chern-Simons fractional fluxes where the flux superpotential
could be small enough to compete with gaugino condensation, even if this case would lead to supersymmetry breaking around the GUT scale;

\item Compactifications on non-K\"ahler manifolds which do not admit a closed holomorphic $(3,0)$-form,
since these cases allow for new geometric fluxes which could play a similar r\^ole as type IIB Ramond-Ramond fluxes,
and could be used to tune $W_0$ to small values \cite{Lopes Cardoso:2002hd,CCDL, Held:2010az, Becker:2003gq, Becker:2003yv}.
In this case one could lower the gravitino mass to the TeV scale and have enough freedom to tune the cosmological constant.
However, the moduli space of these manifolds is at present only poorly understood.
\end{enumerate}
In this paper, we shall not consider the second option given that we want to focus on cases, like `special Hermitian manifolds',
which represent the smallest departure from a CY due to $\alpha'$ effects.
\end{itemize}

This analysis suggests that if one is interested in deriving vacua
where our K\"ahler moduli stabilisation mechanism leads to spontaneous
supersymmetry breaking around the TeV scale, one should focus on one of the two following situations:
\begin{enumerate}
\item Models where the requirement of a holomorphic gauge bundle fixes all complex structure moduli
without inducing singularities (so that the supergravity approximation is reliable), i.e. models with $h^{1,2}_{\rm hol}=0$ \cite{AGLO1,AGLO2,Anderson:2011ty}.
The dilaton could then be fixed in a supersymmetric way by using a double gaugino condensate while the K\"ahler
moduli could be fixed following our LVS-like method by including worldsheet instantons, threshold and $\alpha'$ effects.
This global minimum would break supersymmetry spontaneously along the K\"ahler moduli directions.
The gravitino mass could then be around the TeV scale because of the exponential suppression from gaugino condensation.

\item Simple models with a very small number of complex structure moduli, like Abelian orbifolds with a few untwisted $Z$-moduli,
or even non-Abelian orbifolds with no complex structure moduli at all. In fact, in this case gauge singlets
could be fixed at non-zero values via D-terms induced by anomalous $U(1)$ factors and higher order terms in the superpotential \cite{Buchmuller:2005jr,Buchmuller:2006ik,Lebedev:2006kn,Lebedev:2007hv,Lebedev:2008un},
so resulting in cases where all the $Z$-moduli become massive by the holomorphicity of the gauge bundle.
The dilaton could then be fixed by balancing gaugino condensation with the
contribution from a gauge bundle modulus (i.e. a continuous Wilson line in the orbifold language)
which develops a small vacuum expectation value (VEV)
because it comes from $R$-symmetry breaking higher order terms in the superpotential \cite{HPN,Kappl:2008ie}.
A low gravitino mass could then be obtained due to this small VEV.
\end{enumerate}
Let us finally note that accommodating our observed cosmological
constant, which is a challenge even with fluxes and $\mc{O}(100)$ complex structures,
is even more of a challenge in cases without quantised fluxes.

This paper is organised as follows. In Section~\ref{sec:framework} we introduce the general framework of heterotic CY compactifications \cite{Witten:1985xb,Ferrara:1986qn}, reviewing the form of the tree-level effective action and then presenting
a systematic discussion of quantum corrections from non-perturbative effects~\cite{Font:1990nt,Ferrara:1990ei,Nilles:1990jv}, string loops~\cite{BHK,BHP,CCQ}, and higher-derivative $\alpha'$-corrections~\cite{AQS,StandardAlphaPrime,BBHL} according to their successive level of suppression
by powers of the string coupling and inverse powers of the volume.
Supersymmetric vacua are then discussed in Section~\ref{SusyVacua},
while in Section~\ref{NonSusyVacua} we derive new global minima with spontaneous supersymmetry breaking which can
even be Minkowski (or slightly de Sitter) if enough tuning freedom is available.
After discussing in Section~\ref{sec:Softmass} the resulting pattern of moduli and soft masses generated by gravity mediation,
we derive anisotropic constructions in Section~\ref{sec:Anisotropic}.
We finally present our conclusions in Section \ref{Conclusions}.

\section{Heterotic framework}
\label{sec:framework}

Let us focus on weakly coupled heterotic string theory compactified on a smooth CY three-fold $X$.
The 4D effective supergravity theory involves several moduli:
$h^{1,2}(X)$ complex structure moduli $Z_{\alpha}$, $\alpha=1,...,h^{1,2}(X)$;
the dilaton $S$ and $h^{1,1}$ K\"ahler moduli $T_i$, $i=1,...,h^{1,1}(X)$ (besides several gauge bundle moduli).

The real part of $S$ is set by the 4D dilaton (see appendix \ref{AppA} for the correct normalisation):
\be
{\rm Re}(S) \equiv s = \frac{1}{4\pi}\,e^{-2\phi_4} = \frac{1}{4\pi}\, e^{-2\phi}\,\vo\,,
\ee
where $\phi$ is the 10D dilaton whose VEV gives the string coupling $e^{\langle\phi\rangle}=g_s$.
The imaginary part of $S$ is given by the universal axion $a$ which is the 4D dual of $B_2$.
On the other hand, the real part of the K\"ahler moduli, $t_i = {\rm Re}(T_i)$, measures the volume of internal two-cycles
in units of the string length $\ell_s=2\pi\sqrt{\alpha'}$. The imaginary part of $T_i$ is
given by the reduction of $B_2$ along the basis $(1,1)$-form $\hat{D}_i$ dual to the divisor $D_i$.

We shall focus on general non-standard embeddings with possible $U(1)$ factors in the visible sector.
Hence the gauge bundle in the visible $E_8^{{\rm vis}}$ takes the form $V_{{\rm vis}}=U_{\rm vis} \bigoplus_{\kappa} \mc{L}_{\kappa}$
where $U_{{\rm vis}}$ is a non-Abelian bundle whereas the $\mc{L}_{\kappa}$ are line bundles. On the other
hand the vector bundle in the hidden $E_8^{{\rm hid}}$ involves just a non-Abelian factor
$V_{{\rm hid}}=U_{\rm hid}$. We shall not allow line bundles in the hidden sector
since, just for simplicity, we shall not consider matter fields charged under anomalous $U(1)$s.
In fact, if we want to generate a superpotential from gaugino condensation
in the hidden sector in order to fix the moduli, all the anomalous $U(1)$s have to reside in the
visible sector otherwise, as we shall explain later on, the superpotential would not be gauge invariant.

\subsection{Tree-level expressions}
\label{TLexp}

The tree-level K\"ahler potential takes the form:
\be
K_{\rm tree}= - \ln\vo-\ln(S+\overline{S}) -\ln\left({\rm i}\int_X\Omega\wedge \overline{\Omega} \right),
\label{Ktree}
\ee
where $\vo$ is the CY volume measured in string units, while $\Omega$ is the
holomorphic $(3,0)$-form of $X$ that depends implicitly on the $Z$-moduli. The internal volume
depends on the $T$-moduli since it looks like:
\be
\vo=\frac 16 \,k_{ijk} t_i t_j t_k= \frac{1}{48} \,k_{ijk} \left(T_i+\overline{T}_i\right) \left(T_j+\overline{T}_j\right) \left(T_k+\overline{T}_k\right),
\ee
where $k_{ijk}=\int_X \hat{D}_i\wedge \hat{D}_j \wedge\hat{D}_k$ are the triple intersection numbers of $X$.

The tree-level holomorphic gauge kinetic function for both the visible and hidden sector is given
by the dilaton:
\be
f_{\rm tree} = S \qquad \Rightarrow \qquad {\rm Re}(f_{\rm tree})\equiv g^{-2}_{\scriptscriptstyle 4} = s\,.
\label{ftree}
\ee
The tree-level superpotential is generated by the three-form flux $H$ and it reads:
\be
W_{\rm flux} = \int_X H\wedge \Omega\,,
\label{Wtree}
\ee
with the correct definition of $H$ including $\alpha'$ effects:
\be
H= d B_2 - \frac{\alpha'}{4}\left[{\rm CS}(A)-{\rm CS}(\omega)\right],
\ee
where ${\rm CS}(A)$ is the Chern-Simons three-form for the gauge connection $A$:
\be
{\rm CS}(A) = {\rm Tr} \left(A\wedge dA +\frac 23\, A\wedge A\wedge A\right),
\ee
and ${\rm CS}(\omega)$ is the gravitational equivalent for the spin connection $\omega$.

The VEV of the tree-level superpotential, $W_0$, is of crucial importance.
Due to the difference with type IIB where one has two three-form fluxes,
which can give rise to cancellations among themselves leading to small values of $W_0$,
in the heterotic case $W_0$ is generically of order unity. Hence one experiences two problems:
\begin{enumerate}
\item Contrary to type IIB, the heterotic dilaton is not fixed by the flux superpotential,
resulting in a supergravity theory which is not of no-scale type. More precisely,
the F-term scalar potential:
\be
V_F= e^K \left(K^{I\bar{J}}D_I W D_{\bar{J}}\bar{W} -3 |W|^2\right),
\ee
derived from (\ref{Ktree}) and (\ref{Wtree}) simplifies to:
\bea
V_F&=& e^K \left[\sum_{Z}K^{\alpha\bar{\beta}}D_{\alpha} W D_{\bar{\beta}}\bar{W}
+\left(K^{S\bar{S}} K_S K_{\bar{S}} +\sum_{T}K^{i\bar{j}} K_i K_{\bar{j}} -3\right) |W|^2\right] \nn \\
&& = e^K \left(\sum_{Z}K^{\alpha\bar{\beta}}D_{\alpha} W D_{\bar{\beta}}\bar{W}
+|W|^2\right),
\label{Vtree}
\eea
since $K^{S\bar{S}} K_S K_{\bar{S}} =1$ and $\sum_{T}K^{i\bar{j}} K_i K_{\bar{j}}-3=0$. Setting $D_{\alpha} W=0$
$\forall \alpha=1,...,h^{1,2}(X)$, the scalar potential (\ref{Vtree}) reduces to:
\be
V_F = e^K |W_0|^2 = \frac{|W_0|^2}{2s \vo}\,,
\ee
yielding a run-away for both $s$ and $\vo$ if $|W_0|\neq 0$.
Given that generically $|W_0|\simeq \mc{O}(1)$, it is very hard to balance this tree-level run-away
against $S$-dependent non-perturbative effects which are exponentially suppressed in $S$.
One could try to do it by considering small values of $s = g^{-2}_{\scriptscriptstyle 4D}$
but this would involve a strong coupling limit where control over moduli stabilisation is lost.
A possible way to lower $|W_0|$ was proposed in \cite{GKLM} where the authors derived the topological
conditions to have fractional Chern-Simons invariants.

\item If $|W_0|\neq 0$, even if it is fractional, one cannot obtain low-energy supersymmetry. In fact,
the gravitino mass is given by $m_{3/2} = e^{K/2} |W_0| M_P$, and so the invariant quantity $e^{K/2} |W_0| = |W_0|/(\sqrt{2s\vo})$
has to be of order $10^{-15}$ to have TeV-scale supersymmetry. As we have seen, the 4D gauge coupling is
given by $\alpha_\GUT^{-1} = g_s^{-2}\vo$, and so
a huge value of the internal volume would lead to a hyper-weak GUT coupling. Note that a very large value of
$\vo$ cannot be compensated by a very small value of $g^{-2}_s$ since we do not want to
violate string perturbation theory.
\end{enumerate}

Let us briefly mention that in some particular cases one could have an
accidental cancellation among the flux quanta which yields a small $|W_0|$ as suggested in \cite{BdA}.
We stress that in the heterotic case, contrary to type IIB, this cancellation
is highly non-generic, and so it is not very appealing to rely on it to lower $|W_0|$.
Hence it would seem that the most promising way to get low-energy supersymmetry is to
consider the case where $|W_0|=0$ and generate an exponentially small superpotential only at sub-leading non-perturbative level.
This case was considered in \cite{AGLO2}, where the authors argued that, at tree-level,
one can in principle obtain a Minkowski supersymmetric vacuum
with all complex structure moduli stabilised and $2(h^{1,1}+1)$
flat directions corresponding to the dilaton and the K\"ahler moduli.
As explained in Section \ref{Introduction}, this corresponds to the best case scenario
where the gauge bundle is holomorphic only at a non-singular point-like sub-locus in complex structure moduli space.

If instead one focuses on the more general case where $h^{1,2}_{\rm hol}>0$ $Z$-moduli are left flat
after imposing the requirement of a holomorphic gauge bundle,
as we shall show in section \ref{SusyVacua}, the conditions $D_{Z^\alpha} W_{\rm flux}=0$
$\forall \alpha=1,...,h^{1,2}_{\rm hol}$ and $|W_0|=0$ imply that no quantised $H$ flux is turned on,
resulting in the impossibility to stabilise the remaining $Z$-moduli.
This result implies that it is impossible to stabilise the remaining complex structure moduli and the dilaton in two steps
with a $Z$-moduli stabilisation at tree-level and a dilaton stabilisation at sub-leading non-perturbative level.
In this case there are two possible way-outs:
\begin{enumerate}
\item Focus on the case $D_{Z^\alpha} W=0$
$\forall \alpha=1,...,h^{1,2}_{\rm hol}$ and $|W_0|\neq 0$ so that $H$ can be non-trivial. In this case one has however a
dilaton run-away, implying that no moduli can be fixed at tree-level. One needs therefore to add $S$-dependent non-perturbative
effects which have to be balanced against the tree-level superpotential to lift the run-away. A small $|W_0|$ could be obtained
either considering fractional Chern-Simons invariants or advocating accidental cancellations among the flux quanta.

\item Focus on the case with trivial $H$ so that no scalar potential is generated at tree-level.
The dilaton and the complex structure moduli could then be fixed at non-perturbative level via a race-track superpotential
generating an exponentially small $W$ which could lead to low-energy supersymmetry.
Note that even though $dH=R\wedge R-F\wedge F\neq 0$ for $(0,2)$-models, it is still possible to have $|W_0|=0$
since only the harmonic part of the $H$-flux contributes to this superpotential (see discussion in section \ref{sec:SUSYvac}).
Hence, moduli stabilisation would have to proceed via a racetrack mechanism involving at least two condensing gauge groups with {\it all} moduli appearing in the gauge kinetic functions and/or the prefactors of the non-perturbative terms. Since this is generically not the case for heterotic compactifications, this avenue will not lead to supersymmetric moduli stabilisation except perhaps for a few specific cases. Note that in this case to get a Minkowski supersymmetric vacuum one would have to fine-tune the prefactors of the two (or more) condensates so that $W=0$ at the minimum. Then one would have (under the conditions mentioned above) a set of holomorphic equations for the $Z$-moduli which will always have a solution. However once supersymmetry is broken this option is no-longer available since now one needs to have $W\ne 0$ at the minimum if one is to have any hope of fine-tuning the cosmological constant to zero. However now the equations for the $Z$-moduli are a set of real non-linear equations which are not guaranteed to have a solution.
\end{enumerate}

\subsection{Corrections beyond leading order}

As explained in the previous section, in smooth heterotic compactifications with $h^{1,2}_{\rm hol}>0$
complex structure moduli not fixed by the holomorphicity of the gauge bundle,
these $Z$-moduli cannot be frozen at tree-level by turning on a quantised background flux since
this stabilisation would need $|W_0|\neq 0$ which, in turn, would induce a dilaton and volume runaway.
Thus, one has to look at any possible correction beyond the leading order expressions.
Before presenting a brief summary of the various effects to be taken into account
(perturbative and non-perturbative in both $\alpha'$ and $g_s$), let us mention two well-known control issues
in heterotic constructions:
\begin{itemize}
\item \emph{Tension between weak coupling and large volume}:
In order to have full control over the effective field theory,
one would like to stabilise the moduli in a region of field space where
both perturbative and higher derivative corrections are small, i.e. respectively for $g_s\ll 1$
and $\vo \gg 1$.
However, as we have already pointed out, this can be the case only if the 4D coupling
is hyper-weak, in contrast with phenomenological observations.
In fact, we have:
\be
\frac{g_s^2}{\vo} = \alpha_\GUT\simeq \frac{1}{25}\,,
\ee
and so if we require $g_s\lesssim 1$, the CY volume
cannot be very large, $\vo\lesssim 25$, implying that one has never a solid parametric control over the approximations
used to fix the moduli.

\item \emph{Tension between GUT scale and large volume}:
In heterotic constructions, the unification scale is identified with the Kaluza-Klein scale, $M_\GUT=M_\KK$,
which cannot be lowered that much below the string scale for $\vo\lesssim 25$, resulting in a GUT scale which is
generically higher than the value inferred from the 1-loop running of the MSSM gauge couplings.
In more detail, the string scale $M_s\equiv\ell_s^{-1}$ can be expressed in terms of the 4D Planck scale from dimensional reduction
as (see appendix \ref{AppA} for an explicit derivation):
\be
M_s^2 = \frac{M_P^2}{4\pi\alpha_\GUT^{-1}} \simeq \frac{M_P^2}{100\pi}\simeq \left(1.35\cdot 10^{17}\,\text{GeV}\right)^2\,.
\ee

In the case of an isotropic compactification, the Kaluza-Klein scale takes the form:
\be
M_\GUT=M_\KK \simeq \frac{M_s}{\vo^{1/6}}\gtrsim 8\cdot 10^{16}\,{\rm GeV}\qquad\text{for}\qquad\vo\lesssim 25\,,
\ee
which is clearly above the phenomenological value $M_\GUT\simeq 2.1 \cdot 10^{16}$ GeV.
On the other hand, anisotropic compactifications with $d$ large dimensions of size $L=x\ell_s$ with $x\gg 1$
and $(6-d)$ small dimensions of string size $l=\ell_s$, can lower the Kaluza-Klein scale:
\be
{\rm Vol}(X) = L^d l^{6-d} = x^d \ell_s^6=\vo \,\ell_s^6\quad\Rightarrow\quad M_\GUT=M_\KK \simeq \frac{M_s}{x}\simeq \frac{M_s}{\vo^{1/d}}\,.
\ee
For the case $d=2$, one would get the encouraging result $M_\GUT=\frac{M_s}{\sqrt{\vo}} \gtrsim 2.7\cdot 10^{16}$ GeV.
\end{itemize}

\subsubsection{Higher derivative effects}
\label{AlphaPrime}

Let us start considering higher derivative effects, i.e. perturbative $\alpha'$ corrections to the K\"ahler potential.
In the case of the standard embedding corresponding to $(2,2)$ worldsheet theories, the leading $\alpha'$ correction
arises at $\mc{O}(\alpha'^3)\mc{R}^4$ \cite{StandardAlphaPrime} and depends on the CY Euler number $\chi(X)= 2\left(h^{1,1}-h^{1,2}\right)$.
Its form can be derived by substituting the $\alpha'$ corrected volume $\vo \to \vo +\xi/2$ into the tree-level expression (\ref{Ktree})
with $\xi= - \zeta(3)\chi(X)/(2 (2\pi)^3)$. Given that $\zeta(3)\simeq 1.2$, $\xi$ is of the order $\xi\simeq \left(h^{1,2}-h^{1,1}\right)/200\simeq \mc{O}(1)$
for ordinary CY three-folds with $\left(h^{1,2}-h^{1,1}\right)\simeq \mc{O}(100)$. Hence for $\vo\simeq \mc{O}(20)$, the
ratio $\xi/(2\vo)\simeq \mc{O}(1/40)$ is a small number which justifies the expansion:
\be
K \simeq -\ln\vo - \frac{\xi}{2\vo}\qquad\Rightarrow\qquad K_{\alpha'^3}=-\frac{\xi}{2\vo}\,.
\label{Kalpha'3}
\ee
As pointed out in \cite{AQS} however, this is the leading order higher derivative effect only for the standard embedding
since $(0,2)$ worldsheet theories admit $\alpha'$ corrections already at $\mc{O}(\alpha'^2)$ which deform the K\"ahler form $J$ as:
\be
J \to J' = J +\mc{O}(\alpha') \,\tilde{h} +\mc{O}(\alpha'^2) \,\tilde{h}^{(2)} + ...\,,
\label{Jdef}
\ee
where both $\tilde{h}$ and $\tilde{h}^{(2)}$ are moduli-dependent $(1,1)$-forms which are orthogonal to $J$,
i.e. $\int_X \ast J \wedge  \tilde{h} = \int_X \ast J \wedge  \tilde{h}^{(2)}=0$. Plugging $J'$ into the
tree-level expression for $K$ (\ref{Ktree}) and then expanding, one finds that the $\mc{O}(\alpha')$ correction
vanishes because of the orthogonality between $\tilde{h}$ and $J$ whereas at $\mc{O}(\alpha'^2)$ one finds:\footnote{In looking at the derivation of the correction at $\mc{O}(\alpha'^2)$ in~\cite{AQS}, one may wonder about the r\^ole of field redefinitions. The fact that the corrected K\"ahler potential $K'$ can be written in terms of $J'$ as a function of $\int J'\wedge J'\wedge J'$ alone, just the same way as the tree-level $K$ in terms of $J$, may imply that a field redefinition of the K\"ahler form may actually fully absorb the correction at ${\cal O}(\alpha'^2)$. To this end, the observation in~\cite{AQS} that the generically non-vanishing string 1-loop corrections in type IIB appearing at ${\cal O}(\alpha'^2)$ are S-dual to the heterotic correction, provides additional evidence for the existence of this term.}
\be
K_{\alpha'^2} = \frac{1}{2\vo}\int_X \ast \tilde{h}\wedge \tilde{h} = \frac{||\tilde{h}||^2}{2\vo}\,.
\label{Kalpha2}
\ee
Note that the correction (\ref{Kalpha2}) is generically leading with respect to (\ref{Kalpha'3})
since (\ref{Kalpha2}) should be more correctly rewritten as:
\be
K_{\alpha'^2} = \frac{g}{\vo^{2/3}}\qquad\text{with}\qquad g\equiv\frac{||\tilde{h}||^2}{2\vo^{1/3}}=-\frac{1}{2\vo^{1/3}}\int_X J\wedge \tilde{h}\wedge \tilde{h}\geq 0\,,
\label{Kalpha'2}
\ee
where $g$ is a homogeneous function of the K\"ahler moduli of degree 0 given that $J$ scales as $J\sim \vo^{1/3}$
and $\tilde{h}$ does not depend on $\vo$. As an illustrative example,
let us consider the simplest Swiss-cheese CY $X$ with one large two-cycle $t_b$ and one small blow-up
mode $t_s$ so that $J = t_b \hat{D}_b - t_s \hat{D}_s$ and the volume reads:
\be
\vo = k_b t_b^3 - k_s t_s^3 >0\qquad\text{for}\qquad 0\leq \frac{t_s}{t_b}< \left(\frac{k_b}{k_s}\right)^{1/3}\,.
\ee
In the limit $k_b t_b^3\gg k_s t_s^3$, the function $g$ then becomes (considering, without loss of generality, $\tilde{h}$ as moduli-independent):
\be
g = c_b + c_s \,\frac{t_s}{t_b}\geq 0\qquad\text{with}\quad c_b = -\frac{1}{2\,k_b^{1/3}} \int_X \hat{D}_b \wedge \tilde{h}\wedge \tilde{h}
\quad\text{and}\quad c_s = \frac{1}{2\,k_b^{1/3}} \int_X \hat{D}_s \wedge \tilde{h}\wedge \tilde{h}\,. \label{gfun}
\ee
The sign of $c_b$ and $c_s$ can be constrained as follows. In the limit $t_s/t_b \to 0$, $g$ reduces to $g =c_b = |c_b|\geq 0$.
On the other hand, requiring that $g$ is semi-positive definite for any point in K\"ahler moduli space one finds:
\be
c_s = - \,|c_b|\,\left(\frac{k_s}{k_b}\right)^{1/3} + |\kappa|\,, \label{cs}
\ee
where $|\kappa|$ is a semi-positive definite quantity.

\subsubsection{Loop effects}

Let us now focus on $g_s$ perturbative effects which can modify both the K\"ahler potential and the
gauge kinetic function. The exact expression of the string loop corrections to the K\"ahler potential
is not known due to the difficulty in computing string scattering amplitudes on CY backgrounds.
However, in the case of type IIB compactifications, these corrections have been argued to be sub-leading compared to $\alpha'$ effects
by considering the results for simple toroidal orientifolds \cite{BHK} and trying to generalise them to arbitrary CY backgrounds \cite{BHP,CCQ}.
Following \cite{CCQ}, we shall try to estimate the behaviour of string loop corrections to the scalar potential by demanding that these
match the Coleman-Weinberg potential:
\be
V_{g_s}\simeq \Lambda^2 \,{\rm Str}\, M^2\simeq m_{3/2}^2 \,M_\KK^2 \simeq \frac{|W|^2}{2s}  \frac{M_P^4}{\vo^{2(1+1/d)}}\,,
\label{Vgs}
\ee
where we took the cut-off scale $\Lambda= M_\KK$ and we considered $d$ arbitrary large dimensions.
Note that these effects are indeed subdominant with respect to the $\alpha'$ ones for large volume
since the $\mc{O}(\alpha'^2)$ and $\mc{O}(\alpha'^3)$ corrections, (\ref{Kalpha'2}) and (\ref{Kalpha'3}), give
respectively a contribution to the scalar potential of the order $V_{\alpha'^2}\simeq |W|^2 /\vo^{5/3}$
and $V_{\alpha'^3}\simeq |W|^2/\vo^2$, whereas the $g_s$ potential (\ref{Vgs}) scales as $V_{g_s}\simeq |W|^2 /\vo^{7/3}$
for the isotropic case with $d=6$ and $V_{g_s}\simeq |W|^2 /\vo^3$ for the anisotropic case with $d=2$.
Due to this subdominant behaviour of the string loop effects, we shall neglect them in what follows.

String loops correct also the gauge kinetic function (\ref{ftree}). The 1-loop correction has a different expression
for the visible and hidden $E_8$ sectors \cite{BHW}:
\be
f_{\rm vis} = S + \frac{\beta_i}{2}\, T_i\,, \qquad f_{\rm hid} = S - \frac{\beta_i}{2}\, T_i\,,
\ee
where:
\be
\beta_i = \frac{1}{4\pi}\int_X \left(c_2 (V_{\rm vis}) - c_2(V_{\rm hid})\right) \wedge \hat{D}_i\,.
\ee

\subsubsection{Non-perturbative effects}

The 4D effective action receives also non-perturbative corrections in both $\alpha'$ and $g_s$.
The $\alpha'$ effects are worldsheet instantons wrapping an internal two-cycle $T_i$. These
give a contribution to the superpotential of the form:
\be
W_{\rm wi} = \sum_j B_j \,e^{-\,b_{ij} T_i}\,.
\label{Wwi}
\ee
Note that these contributions arise only for $(0,2)$ worldsheet theories whereas they
are absent in the case of the standard embedding.
On the other hand, $g_s$ non-perturbative effects include gaugino condensation and NS5 instantons.
In the case of gaugino condensation in the hidden sector group, the resulting superpotential looks like:
\be
W_{\rm gc} = \sum_j A_j\, e^{-\,a_j \,f_{\rm hid}} = \sum_j A_j\,e^{-\,a_j \left(S - \frac{\beta_i}{2} T_i\right)}\,,
\label{Wgc}
\ee
where in the absence of hidden sector $U(1)$ factors, all the hidden sector gauge groups have the same gauge kinetic
function. Finally, NS5 instantons wrapping the whole CY manifold would give a sub-leading non-perturbative
superpotential suppressed by $e^{-\vo}\ll 1$, and so we shall neglect them.

\subsection{Moduli-dependent Fayet-Iliopoulos terms}

As already pointed out, we shall allow line bundles in the visible sector where
we turn on a vector bundle of the form $V_{{\rm vis}}=U_{\rm vis} \bigoplus_{\kappa} \mc{L}_{\kappa}$.
The presence of anomalous $U(1)$ factors induces $U(1)$ charges for the moduli
in order to cancel the anomalies and gives rise to moduli-dependent Fayet-Iliopoulos (FI) terms.
In particular, the charges of the K\"ahler moduli and the dilaton under the $\kappa$-th anomalous $U(1)$ read:
\be
q^{(\kappa)}_{T_i}=4\, c_1^i(\mc{L}_{\kappa})\qquad \text{and}\qquad q_s^{(\kappa)}= 2\,\gamma_{(\kappa)}=2\,\beta_i \,c_1^i(\mc{L}_{\kappa})\,,
\label{U(1)charges}
\ee
so that the FI-terms become \cite{BHW}:
\be
\xi_{(\kappa)}= - q^{(\kappa)}_{T_i} \frac{\partial K}{\partial T_i} - q_s^{(\kappa)}\frac{\partial K}{\partial S}=
\frac{c_1^i(\mc{L}_{\kappa})}{\vo} \,k_{ijk} t_j t_k+\frac{\gamma_{(\kappa)}}{s}\,.
\label{FIterms}
\ee
Note that the dilaton-dependent term in the previous expression is a 1-loop correction to the FI-terms
which at tree-level depend just on the K\"ahler moduli. The final D-term potential takes the form:
\be
V_D = \sum_\kappa \frac{\xi_{(\kappa)}^2}{{\rm Re}\left(f_{(\kappa)}\right)}\,.
\label{VD}
\ee
From the expressions (\ref{U(1)charges}) for the $U(1)$-charges of the moduli, we can now check the $U(1)$-invariance
of the non-perturbative superpotentials (\ref{Wwi}) and (\ref{Wgc}). In the absence of charged matter fields,
the only way to obtain a gauge invariant worldsheet instanton is to choose the gauge bundle such that all the $T_i$
appearing in $W_{\rm wi}$ are not charged, i.e. $c_1^i(\mc{L}_{\kappa})=0$ $\forall \kappa$ and $\forall i$.
The superpotential generated by gaugino condensation is instead automatically $U(1)$-invariant by construction since
all the anomalous $U(1)$s are in the visible sector whereas gaugino condensation takes place in the hidden sector.
Thus, the hidden sector gauge kinetic function is not charged under any anomalous $U(1)$:
\be
q_{f_{\rm hid}}^{(\kappa)} = q_s^{(\kappa)}-\frac{\beta_i}{2}\,q_{T_i}^{(\kappa)}
= 2 \left(\gamma_{(\kappa)} - \beta_i c_1^i(\mc{L}_{\kappa})\right)=0\,.
\ee
Before concluding this section, we recall that in supergravity the D-terms are proportional to the F-terms for $W\neq 0$.
In fact, the total $U(1)$-charge of the superpotential $W$ is given by $q^{(\kappa)}_W = q^{(\kappa)}_i W_i/W = 0$, and so one can write:
\be
\xi_{(\kappa)} = -\,q_i^{(\kappa)} K_i = -\,q_i^{(\kappa)} \frac{D_i W}{W} = -\,q_i^{(\kappa)} \frac{e^{-K/2}}{W}\,K_{i\bar{j}}\bar{F}^{\bar{j}}\,,
\label{FDrelation}
\ee
where the F-terms are defined as $F^i=e^{K/2}K^{i\bar{j}}D_{\bar{j}}\bar{W}$. Therefore if all the F-terms are vanishing with $W\neq 0$,
the FI-terms are also all automatically zero without giving rise to independent moduli-fixing relations.

\section{Supersymmetric vacua}
\label{SusyVacua}

In this section, we shall perform a systematic discussion of heterotic supersymmetric vacua
starting from an analysis of the tree-level scalar potential and then including corrections beyond the leading order expressions.

\subsection{Tree-level scalar potential}
\label{sec:SUSYvac}

In \cite{Strominger:1986uh}, Strominger analysed the 10D equations of motion and worked out the
necessary and sufficient conditions to obtain $\mc{N}=1$ supersymmetry in 4D assuming a 10D space-time of the form
$M \times X$ where $M$ is a maximally symmetric 4D space-time and $X$ is a compact 6D manifold:
\begin{enumerate}
\item $M$ is Minkowski;

\item $X$ is a complex manifold, i.e. the Nijenhuis tensor has to vanish;

\item There exists one globally defined holomorphic $(3,0)$-form $\Omega$ which is closed, i.e. $d\Omega=0$, and whose norm is related to the complex structure $(1,1)$-form $J$ as (up to a constant):\footnote{The adjoint operator $d^\dagger$ can be defined from the inner product $\langle \omega, \sigma \rangle = \int_X \omega \wedge \ast \sigma$ as $\langle \omega_p , d \omega_{p-1}\rangle = \langle d^\dagger \omega_p, \omega_{p-1}\rangle$. For an even dimensional manifold, as we have here, $d^\dagger =-\ast d\ast$.}
\be
d^\dagger J =i (\partial-\bar\partial) \ln ||\Omega||
\ee

\item The background gauge field $F$ has to satisfy the Hermitian Yang-Mills equations:
\be
F_{(0,2)}=F_{(2,0)}=0 \qquad \text{and} \qquad g^{i\bar{j}}F_{i\bar{j}}=0
\label{HYMeqs}
\ee

\item The dilaton $\phi$ and the warp factor $A$ have to satisfy (again up to a constant):\footnote{We are writing the total metric
as $ds^2 = e^{2 A(y)}\left(g_{\mu\nu}(x)dx^\mu dx^\nu + g_{ij}(y) dy^i dy^j\right)$.}
\be
\phi(y)= A(y)=\frac 18 \ln ||\Omega||(y)
\ee

\item The background three-form flux is given by:
\be
H=-\frac{i}{2}(\partial-\bar{\partial})J\,,
\label{Scon}
\ee
together with the Bianchi identity:
\be
dH=-\frac{\alpha'}{4}\left[{\rm tr}(F\wedge F)-{\rm tr}(R\wedge R)\right].
\label{Bianchi}
\ee
\end{enumerate}

Some of the conditions listed above can be reformulated also in terms of constraints on the five torsional classes $\mc{W}_i$, $i=1,...,5$ (for a review see \cite{Grana:2005jc,Lopes Cardoso:2002hd}). The second condition corresponds to $\mc{W}_1=\mc{W}_2=0$ implying that the torsional class $\tau$ belongs to the space
$\tau \in \mc{W}_3\oplus\mc{W}_4\oplus\mc{W}_5$. This is the case of `Hermitian manifolds'. Moreover, the third condition above
gives $\mc{W}_5=-2\mc{W}_4 = d\ln||\Omega||$ implying that both $\mc{W}_4$ and $\mc{W}_5$ are exact real 1-forms.
We shall focus on the simplest solution to $2\mc{W}_4+\mc{W}_5=0$ which is $\mc{W}_4=\mc{W}_5=0$ corresponding to the case of `special-Hermitian manifolds'
where the dilaton and the warp factor are constant~\cite{Lopes Cardoso:2002hd}. More general solutions involve a non-constant dilaton profile in the extra dimensions
and $\mc{W}_i\neq 0$ for $i=4,5$ but we shall not consider this option~\cite{Lopes Cardoso:2002hd}.

Let us comment on the implications of the last Strominger condition (\ref{Scon}) which for constant dilaton
can be rewritten as $H=-\frac 12 \ast dJ$. Using the Hodge decomposition theorem, the three-form $H$ can be expanded uniquely as:
\be
H = H_{\rm harm} + H_{\rm exact} + H_{\rm co-exact}\,,
\label{expantion}
\ee
where $H_{\rm harm}$ is a harmonic form, $H_{\rm exact}$ is an exact form and
$H_{\rm co-exact}$ is a co-exact form which are all orthogonal to each other.
Given that $\ast d J = - d^\dagger\ast J$, (\ref{Scon}) implies
that $H$ is a co-exact form, and so $H_{\rm harm} = H_{\rm exact} = 0$. Moreover, since $dJ$
is a $(2,1)+(1,2)$ form, (\ref{Scon}) implies that the $(3,0)+(0,3)$ component of $H_{\rm co-exact}$ is zero while the
$(2,1)+(1,2)$ component breaks the K\"ahler condition $dJ=0$. However this happens only at $\mc{O}(\alpha')$.
In fact, the general expression of the $H$-flux is:
\be
H = H_{\rm flux}+ d B_2 - \frac{\alpha'}{4}\left[{\rm CS}(A)-{\rm CS}(\omega)\right],
\label{Hexpression}
\ee
where $H_{\rm flux}$ is a harmonic piece and the combination of Chern-Simons three-forms can also be decomposed as:
\be
\left[{\rm CS}(A)-{\rm CS}(\omega)\right]={\rm CS}_{\rm harm}+{\rm CS}_{\rm exact}+{\rm CS}_{\rm co-exact}\,.
\ee
Comparing the two expressions for $H$, (\ref{expantion}) and (\ref{Hexpression}), we have (due to the uniqueness of the Hodge decomposition):
\be
H_{\rm harm}=H_{\rm flux}-\frac{\alpha'}{4}\,{\rm CS}_{\rm harm}\,,\quad H_{\rm exact}=dB_2-\frac{\alpha'}{4}\,{\rm CS}_{\rm exact}\,,
\quad H_{\rm co-exact}=-\frac{\alpha'}{4}\,{\rm CS}_{\rm co-exact}\,. \nn
\ee
Then the relation (\ref{Scon}) takes the form:
\be
\frac{\alpha'}{2}\,{\rm CS}_{\rm co-exact}=\ast \,dJ\,,
\label{SconNew}
\ee
showing exactly that the K\"ahler condition $dJ=0$ is violated at $O(\alpha')$.
Note that this would be the case for the non-standard embedding where ${\rm CS}_{\rm co-exact}\neq 0$
contrary to the less generic situation of the standard embedding where the Chern-Simons piece vanishes.
Taking the exterior derivative of (\ref{SconNew}) we recover the Bianchi identity (\ref{Bianchi}) which now looks like:
\be
d\ast dJ =\frac{\alpha'}{2}\,\left[{\rm tr}(F\wedge F)-{\rm tr}(R\wedge R)\right].
\ee
This 10D analysis can also be recast in terms of an effective potential
which can be written as a sum of BPS-like terms and whose minimisation reproduces the conditions above \cite{Witten:1985bz,CCDL,Held:2010az,Becker:2003gq}.
Furthermore, some of these conditions can be re-derived as F- or D-term equations of 4D supergravity,
which could lead to the stabilisation of some of the moduli in a Minkowski vacuum.
For example, it has been shown in \cite{Witten:1985bz,Becker:2003gq}, that the second equation in (\ref{HYMeqs})
is equivalent to a D-term condition since:
\be
{\ast_6 1 \cdot}\, g^{i\bar{j}}F_{i\bar{j}} = \frac 12 \,F \wedge J \wedge J\,.
\ee
This D-term condition holds for general non-Abelian gauge fields. If we restrict to Abelian fluxes and integrate the above condition over the CY, this reproduces the tree-level expression for the Fayet-Iliopoulos terms given in (\ref{FIterms}).
If we expand the Abelian fluxes as $F_{(\kappa)} = c_1^i(\mc{L}_{\kappa})\hat{D}_i$ together with $J=t^j\hat{D}_j$ we obtain:
\be
\xi_{(\kappa)} = \frac{1}{\vo}\int_X F_{(\kappa)}\wedge J\wedge J = \frac{c_1^i(\mc{L}_{\kappa})}{\vo} \,k_{ijk} t_j t_k\,,
\ee
which reproduces exactly the tree-level part of (\ref{FIterms}).

Regarding the F-terms, as we have seen in section 2.1, the starting point is the expression of the flux superpotential
which has been inferred in \cite{Becker:2003yv} by comparing the dimensional reduction of the 10D coupling of $H$ to the gravitino mass term in the 4D supergravity action. The final result is:\footnote{In \cite{CCDL} and \cite{Becker:2003gq} it is suggested that
the complete expression for $W$ should more appropriately be
$W=\int_X (H+\frac{i}{2}\,dJ)\wedge\Omega$, similarly to the type IIB case where one has the RR flux in addition
to the $H$-flux. Integrating by parts, the new piece can be rewritten as $\int_X J\wedge d\Omega$ which clearly vanishes since $d\Omega=0$.
However, if one considers the case where $d\Omega\neq 0$, i.e. where supersymmetry is broken directly at the 10D level,
this integral would still be zero if the internal manifold is complex since $dJ$ is of Hodge type $(2,1)+(1,2)$ while $\Omega$ is $(3,0)$.
Thus this term can play a useful r\^ole only for non-complex manifolds with broken supersymmetry.
Due to the difficulty to study this case in a controlled way, we shall not consider it and neglect this additional piece.}
\be
W_{\rm flux}=\int_X H\wedge \Omega = \int_X H_{\rm harm}\wedge \Omega\,.
\label{eq:Wflux}
\ee
Note that only the harmonic component of $H$ contributes to $W_{\rm flux}$.
The harmonic piece $H_{\rm harm}$ can be expanded in a basis of harmonic $(3,0)$- and $(1,2)$-forms as:
\be
H_{\rm harm}= \bar{a}(\bar{Z})\Omega(Z) + b^\alpha(Z,\bar{Z}) \chi_\alpha(Z,\bar{Z}) + c.c.\,,\qquad\alpha=1,...,h^{1,2}(X)\,.
\label{Expan}
\ee
The same $H_{\rm harm}$, together with the holomorphic $(3,0)$-form $\Omega$,
can also be expanded in a symplectic basis of harmonic three-forms ($\alpha_p,\beta^q$) such that $\int_X \alpha_p\wedge \alpha_q=\int_X \beta^p\wedge\beta^q=0$
and $\int_X \alpha_p\wedge\beta^q=\delta_p^q$ with $p,q=0,...,h^{1,2}(X)$:
\be
H_{\rm harm} = e^p \alpha_p-m_q \beta^q\qquad\text{and}\qquad \Omega(Z)=Z^p \alpha_p-G_q(Z) \beta^q\,,
\label{Flux2}
\ee
where $G_q(Z)=\partial_{Z^q} G(Z)$ with $G(Z)$ a homogeneous function of degree 2.
Note that $\alpha_p$ and $\beta^q$ do not depend on the complex structure moduli $Z^\alpha$
which are defined by the expansion of $\Omega$ in (\ref{Flux2}).
If ($A_p,B^q$) is the dual symplectic basis of 3-cycles such that $A_p\cap A_q=B^p\cap B^q=0$
and $A_p\cap B^q=\delta_p^q$, we have (choosing units such that $2\pi\sqrt{\alpha'}=1$):
\be
\int_{B^p} H_{\rm harm} = \int_X H_{\rm harm} \wedge \beta^p
= \int_X \left(e^r \alpha_r-m_q \beta^q\right) \wedge \beta^p= e^p \,,
\label{FluxQuanta2}
\ee
and similarly $m_q=\int_{A_q} H_{\rm harm}$. The quantities $e^p$ and $m_q$ are integer flux quanta.

The expansion of the flux superpotential (\ref{eq:Wflux}) is then given by:
\bea
W_{\rm flux}(Z)&=&\int_X H_{\rm harm}\wedge \Omega = a(Z)\int_X \bar{\Omega}(\bar{Z})\wedge \Omega(Z) \nn \\
&=& {\rm i} a(Z) =m_q Z^q - e^p G_p(Z)\,,
\label{Wfl}
\eea
where we normalised $\int_X \Omega \wedge \bar{\Omega}=- {\rm i}$ and used the fact that $\ast \Omega = -{\rm i}\Omega$ and the orthogonality of the different Hodge components of $H$.

Let us now evaluate the complex structure F-terms $D_{Z^{\alpha}}W_{\rm flux} = \partial_{Z^{\alpha}} W_{\rm flux} + W_{\rm flux} \partial_{Z^{\alpha}} K$.
Using the fact that (see for example \cite{Candelas:1990pi}):
\be
\partial_{Z^\alpha}\Omega=k_{\alpha}(Z,\bar{Z})\Omega+\chi_{\alpha}\,, \qquad
\partial_{Z^\alpha}K = -k_\alpha(Z,\bar{Z})\,, \nn
\ee
and:
\be
K_{\alpha\bar{\beta}}\equiv \partial_{Z^\alpha}\partial_{\bar{Z}^{\bar{\beta}}}K=\frac{\int_X \chi_{\alpha}\wedge \bar{\chi}_{\bar{\beta}}}
{\int_X \Omega\wedge \bar{\Omega}}={\rm i}\int_X \chi_{\alpha}\wedge \bar{\chi}_{\bar{\beta}}\,, \nn
\label{eq:KodSpen}
\ee
and expanding a generic element of the basis of harmonic $(2,1)$-forms as $\chi_\alpha (Z,\bar{Z})= f^p_\alpha(Z,\bar{Z}) \alpha_p- g_{q,\alpha}(Z,\bar{Z}) \beta^q$, we find:
\bea
D_{Z^{\alpha}}W_{\rm flux} &=& \int_X H_{\rm harm} \wedge \chi_{\alpha} = b^{\bar{\beta}}(Z,\bar{Z}) \int_X \bar{\chi}_{\bar{\beta}} \wedge \chi_{\alpha} \nn \\
&=& {\rm i}\,b_{\alpha}(Z,\bar{Z})=m_q f^q_\alpha(Z,\bar{Z}) - e^p g_{p,\alpha}(Z,\bar{Z})\,,
\label{DalphaW}
\eea
where we used again the orthogonality of the different Hodge components of $H$ and the fact that $\ast \chi_\alpha ={\rm i}\chi_\alpha$.
On the other hand, the dilaton and K\"ahler moduli F-terms look like:
\be
D_S W_{\rm flux} = W_{\rm flux} \partial_S K = -{\rm i}\frac{a}{2s}\qquad\text{and}\qquad
D_{T_i} W_{\rm flux} = W_{\rm flux} \partial_{T_i} K = -{\rm i}\frac{a}{4\vo}\,k_{ijk} t_j t_k\,.
\ee
Due to the no-scale cancellation, these F-terms give rise to a scalar potential which is positive definite and reads:
\be
V= e^K \left(\sum_{Z}K^{\alpha\bar{\beta}}D_{\alpha} W D_{\bar{\beta}}\bar{W}+|W|^2\right)
= \frac{1}{2s\vo} \left(\sum_{Z} K^{\alpha\bar{\beta}} b_{\alpha} \bar{b}_{\bar{\beta}} +|a|^2\right)\,.
\label{eq:noscale}
\ee
Let us now set all the F-terms to zero and see what they correspond to:
\begin{itemize}
\item $D_{Z^{\alpha}}W_{\rm flux}=0$ implies that the $(2,1)+(1,2)$ component of $H_{\rm harm}$ is zero.

\item $D_S W_{\rm flux} = 0$ and $D_{T_i} W_{\rm flux}=0$ imply that the $(3,0)+(0,3)$ component of $H_{\rm harm}$
should also be zero, i.e. $W_0\equiv \langle W_{\rm flux}\rangle=0$, if one wants to avoid solutions with a dilaton run-away ($s\to\infty$) or where the internal space decompactifies ($\vo\to\infty$).
\end{itemize}
Combining these two solutions, one has that the total harmonic piece of $H$ should vanish and is of course consistent with
the Strominger condition (\ref{Scon}). An important question to ask now is whether these conditions allow for the fixing
of some moduli. The answer is no. Let us see why.

The first condition $D_{Z^\alpha} W=0$ appears to fix the complex structure moduli supersymmetrically
since one obtains as many equations, $b_\alpha(Z,\bar{Z})=0$, as the number of unknowns (assuming that the $2\,h^{1,2}$
real equations have solutions for some sets of values of the $2\,h^{1,2}+2$ fluxes).
The second condition $W_0\equiv \langle W_{\rm flux}\rangle= {\rm i}\,{a}({Z}) = 0$
could then be satisfied by an appropriate choice of flux quanta.

However the two conditions $D_{Z^\alpha} W=0$ $\Leftrightarrow$ $b_\alpha(Z,\bar{Z})=0$ $\forall \alpha$
and $W_0=0$ $\Leftrightarrow$ $a(Z)=0$ imply from (\ref{Expan}) that $H_{\rm harm}=0$.
Given that $H$ does not depend on the complex structure moduli,
this implies that all flux quanta are zero. In turn, (\ref{Wfl}) and (\ref{DalphaW}) are both identically zero,
and so no potential for the $Z$-moduli is developed. Therefore no moduli, not even the complex structure ones,
can be stabilised at tree-level by using quantised background fluxes.\footnote{The corresponding situation in type IIB is very different since there are two types of fluxes and the effective flux $G_3$ is complex \cite{GKP}.}
In particular, this implies that one cannot perform a two-step stabilisation (similarly to type IIB)
where at tree-level the $Z$-moduli are fixed supersymmetrically
while the $S$- and $T$-moduli are kept flat by tuning $W_0=0$,
and then these remaining moduli are lifted by quantum corrections.
As we have already pointed out in section \ref{TLexp}, we shall avoid this problem by considering
in the next section situations with non-zero flux quanta which allow to fix the $Z$-moduli with $W_0\neq 0$.
The dilaton and volume runaway is then prevented
by scanning over integral and fractional fluxes which give a value of $W_0$
small enough to compete with non-perturbative effects. Hence the system becomes stable only when non-perturbative
effects are included, implying that all the moduli get stabilised beyond tree-level.

\subsubsection{Chern-Simons action and gauge bundle moduli}
\label{Sec311}

In this section, we shall show that also the first equation in (\ref{HYMeqs}), i.e. $F_{(0,2)}=F_{(2,0)}=0$,
can be derived from an F-term condition in 4D supergravity. This requires a brief discussion of
gauge bundle moduli. Let us focus on the Chern-Simons piece of the flux superpotential:
\be
W_{\rm CS}[A]=\int_X {\rm Tr}\left(A\wedge dA +\frac 23\,A\wedge A \wedge A\right) \wedge \Omega\,.
\label{CSsup}
\ee
In the previous expression $A$ is a function of both $x$ and $y$, i.e. non-compact and compact coordinates respectively,
but the differentiation is just $d=dy^{m}\partial_{m}$ since we are only interested in the
contribution to the 4D scalar potential. We shall now write the gauge potential as:
\be
A(x,y) = A_0 (y)+ A_{\rm def} (x,y)\,,
\label{eq:A0A1}
\ee
where $A_0$ is a background contribution independent of $x$ and $A_{\rm def}$ is a generic
deformation which can be parameterised as:
\be
A_{\rm def}(x,y)=\sum_{I=1}^{\infty} C_I(x) \omega^I(y)\,,
\ee
where $C_I$ are 4D scalar fields and $\omega^I$ are an infinite set of $1$-forms living
on $X$ and valued in the adjoint representation of the structure group
of the gauge bundle defined by $A_0$.\footnote{We expect the set of $1$-forms $\omega^I$ to be discrete since
they will be solutions to an elliptic differential equation on the compact manifold.}

\begin{figure}[t]
\begin{center}
\epsfig{file=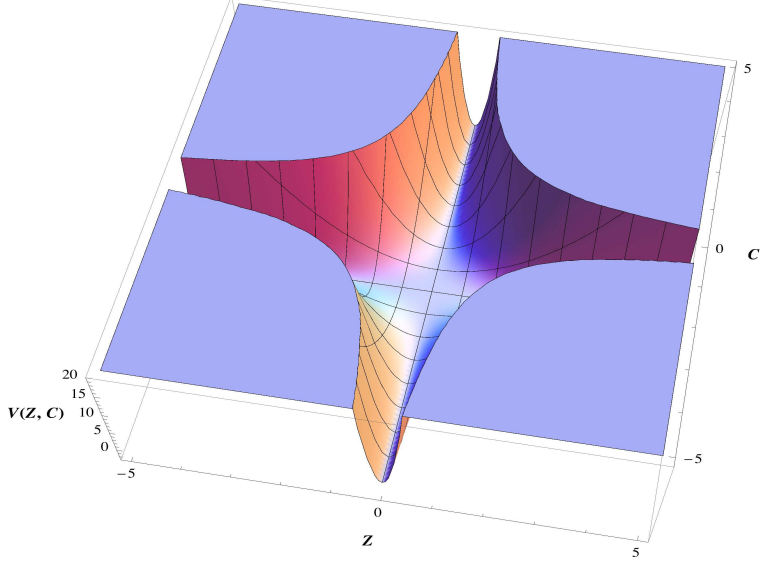, height=75mm,width=100mm}
\caption{Sketch of the leading order scalar potential $V\sim|\partial_C W|^2+\ldots = |Z|^2 |C|^2+\ldots$ as a function of the complex structure moduli (summarily denoted by $Z$) and the gauge bundle moduli (summarily denoted by $C$) as arising at the second order in $W$ schematically as $W=W_{{\rm CS},(2)}C^2\sim (Z+{\cal O}(Z^2))\, C^2$.}
\label{Plot0}
\end{center}
\end{figure}

The superpotential (\ref{CSsup}) then becomes the sum of a constant, a linear, a quadratic and a cubic term in the $C$'s:
\be
W_{\rm CS} = W_{{\rm CS},(0)} + W_{{\rm CS},(1)}^I C_I+ W_{{\rm CS},(2)}^{IJ} C_I C_J + W_{{\rm CS},(3)}^{IJK} C_I C_J C_K\,,
\ee
with (for notational simplicity we dropped the trace symbol):
\bea
W_{{\rm CS},(0)} &=& W_{\rm CS}[A_0]\,, \qquad
W_{{\rm CS},(1)}^I = 2\,\int_X \omega^I\wedge F_0 \wedge \Omega\,, \label{W1} \\
W_{{\rm CS},(2)}^{IJ} &=& \int_X \omega^I \wedge \overline{D}_0\omega^J \wedge \Omega\,, \qquad
W_{{\rm CS},(3)}^{IJK} = \frac 23 \int_X \omega^I \wedge \omega^J \wedge \omega^K \wedge \Omega\,, \label{W2and3}
\eea
where the gauge covariant derivative $\overline{D}_0$ is defined as $\overline{D}_0\omega_{(0,p)}=\overline{\partial} \omega_{(0,p)}
+ A_0 \wedge \omega_{(0,p)} - (-1)^p \omega_{(0,p)}\wedge A_0$
for an arbitrary $(0,p)$-form $\omega_{(0,p)}$. Note that in order to derive these expressions we used $d\Omega=0$,
the anti-commutativity of $d$ and $1$-forms and the cyclicity of the trace.
As we have argued earlier, classically the \textit{total} superpotential
$W$ should be zero at the minimum (for all the moduli), and so the F-term
equation for the bundle moduli $C_I$ is:
\be
0=F_{C_I}=\frac{\partial W_{\rm CS}}{\partial C_I}=W_{{\rm CS},(1)}^I+2 W_{{\rm CS},(2)}^{IJ} C_J +2 W_{{\rm CS},(3)}^{IJK} C_J C_K\,.
\label{eq:FC}
\ee
If $A_{\rm def}$ is a small deformation of the background $A_0$, i.e. $C_I(x) = \varepsilon_I (x)$,
then these F-term equations can be solved order by order in $\varepsilon$, obtaining:
\begin{itemize}
\item At zeroth order $F_{C_I}=0$ gives $W_{{\rm CS},(1)}^I=0$ $\forall I$
which from (\ref{W1}) implies that the $(0,2)$-component of the unperturbed field strength $F_0$ has to vanish.
Hence we recover the holomorphic Yang-Mills equation $F_{0,(0,2)}=0$ which determines (given a complex structure)
$A_0$ to be a flat $(0,1)$ connection. This bundle, which we call $Q_0$, then determines the
exterior derivative operator $\overline{D}_{0}$.

\item At linear order $F_{C_I}=0$ implies (see the expression of $W_{{\rm CS},(2)}^{IJ}$ in (\ref{W2and3})):
\be
W_{{\rm CS},(2)}^{IJ} C_J=0\quad\forall\,I\,\quad\Leftrightarrow \quad C_J\overline{D}_0\omega^J=0\,.
\label{DefMod}
\ee
This equation has two possible solutions:
\begin{enumerate}
\item $\overline{D}_0\omega^i=0\quad \forall\, C_i \quad$ for $i=1,...,N$

\item $\overline{D}_0\omega^{\iota}\neq 0\quad C_{\iota}=0 \quad$ for $\iota=N+1,...,\infty$
\end{enumerate}
The first solution defines the gauge bundle moduli which parameterise all possible deformations of the background that
keep the gauge bundle holomorphic. These first order deformations correspond to $\omega^i\in H^1(End(Q_0))$
where $N\equiv{\rm dim}\left(H^{1}(End(Q_0))\right)$ which is expected to be finite though it may change as one varies the complex structure
since the equations determining the $(0,1)$-forms $\omega^i$ depend on the $Z$-moduli.
Hence $N$ is a function of the $Z$-moduli, i.e. $N=N(Z)$. If $N=0$ for $Z=Z_0$, then
if the complex structure moduli can be stabilised via the fluxes exactly at $Z=Z_0$,
the absence of any gauge bundle moduli is guaranteed (see \cite{Donagi} for similar considerations).
Conversely, the equation $\overline{D}_0\omega^i(Z)=0$ could be used as a mechanism
to reduce the number of complex structure moduli, or even to fix all of them,
if the $C_i$'s develop non-zero VEVs due to D-terms or higher order terms in $W$ \cite{AGLO1,AGLO2,Anderson:2011ty}.
We denoted as $h^{1,2}_{\rm hol}$ the number of $Z$-moduli unconstrained by the equation $\overline{D}_0\omega^i(Z)=0$,
which represents the dimensionality of the sub-locus in complex structure moduli space where the gauge bundle is holomorphic.
In the best case scenario where $h^{1,2}_{\rm hol}=0$ one does not need to turn on any harmonic flux to fix the $Z$-moduli,
whereas in the more general case where $h^{1,2}_{\rm hol}>0$ the remaining complex structure moduli can be fixed
only by turning on a quantised background flux. For a graphical sketch of the `cross-structure' of
the combined complex structure and gauge bundle moduli space see Fig~\ref{Plot0}.

The second solution of (\ref{DefMod}) implies that the forms $\omega^{\iota}$ are not closed under
$\overline{D}_0$ and the index $\iota$ ranges over an infinite set
of values. Hence $C_{\iota}$ are not flat directions but correspond to massive deformations,
namely the Kaluza-Klein modes. We can then easily realise that $W_{{\rm CS},(2)}^{\iota\sigma}$
gives the mass matrix for these Kaluza-Klein modes.

\item Focusing only on the massless modes,
at quadratic order $F_{C_i}=0$ implies:
\be
W_{{\rm CS},(3)}^{ijk} C_j C_k=0\quad\forall i\,,
\label{W3}
\ee
showing that a possible obstruction to the presence of gauge bundle moduli can arise if the Yukawa couplings
are different from zero, i.e. $W_{{\rm CS},(3)}^{ijk}\neq 0$ $\forall i$. We stress again the fact that $W_{{\rm CS},(3)}^{ijk}$ is
a function of the $Z$-moduli, and so even if the equation $\overline{D}_0\omega^i(Z)=0$ (or the flux stabilisation)
gives a solution $Z=Z_\ast$ such that $N(Z_\ast)\neq 0$,
one could still fix all the $C$-moduli if $W_{{\rm CS},(3)}^{ijk}(Z_\ast)\neq 0$ $\forall i$.
\end{itemize}

Having motivated both the background gauge flux and the
nature of the leading deformation we can now work with an arbitrary
deformation by separating the set $\{C_I\}=\{C^{(0)}_i\}\oplus\{C^{\rm KK}_{\iota}\}$
with the first set being the massless modes and the second the Kaluza-Klein
modes. This corresponds to splitting the set of 1-forms as $\{\omega^I\}=\{\omega^i_{(0)}\}\oplus\{\omega^{\iota}_{\rm KK}\}$
where $\overline{D}_0\omega^i_{(0)}=0$ while $\overline{D}_0\omega^\iota_{\rm KK}\neq 0$.
Then under the condition $F_{0,(0,2)}=0$, the F-term equations (\ref{eq:FC}) take the form:
\be
0=F_{C_i^{(0)}}= 2 W_{{\rm CS},(3)}^{ikl} C_k^{(0)} C_l^{(0)}+4 W_{{\rm CS},(3)}^{ik\lambda} C_k^{(0)} C_\lambda^{\rm KK}
+2 W_{{\rm CS},(3)}^{i\sigma\lambda} C_\sigma^{\rm KK} C_\lambda^{\rm KK}\,,
\label{eq:FCi}
\ee
and:
\be
0=F_{C_\iota^{\rm KK}}= 2 W_{{\rm CS},(2)}^{\iota \sigma} C_\sigma^{\rm KK} +2 W_{{\rm CS},(3)}^{\iota kl} C_k^{(0)} C_l^{(0)}
+4 W_{{\rm CS},(3)}^{\iota k\lambda} C_k^{(0)} C_\lambda^{\rm KK}+2 W_{{\rm CS},(3)}^{\iota \sigma\lambda} C_\sigma^{\rm KK} C_\lambda^{\rm KK}\,.
\label{eq:FCiota}
\ee
Note that $W_{{\rm CS},(2)}^{\iota \sigma}\equiv M^{\iota\sigma}_{\rm KK}$ is the mass matrix for the Kaluza-Klein
modes which by definition is non-singular. So eq.~\eqref{eq:FCiota}
can be solved for the massive modes in terms of the massless modes
giving a relation of the form:
\be
C^{\rm KK}_{\sigma}=[M_{\rm KK}]^{-1}_{\sigma\lambda}W_{{\rm CS},(3)}^{\lambda m n} C^{(0)}_m C^{(0)}_n+O(C_0^3)\,.
\ee
Using this in (\ref{eq:FCi}) we get the massless field equation which is the generalisation of (\ref{W3}) for arbitrarily large
deformations of the background gauge bundle:
\be
2 W_{{\rm CS},(3)}^{ikl} C^{(0)}_k C^{(0)}_l +O(C_{(0)}^3)=0\,.
\label{eq:dC00}
\ee
These field equations always admit the solution $C_k^{(0)}=0$ for all gauge bundle moduli
which leaves the complex structure moduli unfixed in the absence of harmonic quantised flux.
Moreover, this solution remains valid even in the presence of non-zero $W$ since the additional term in $D_CW$ is proportional to $C$.
However one could also have solutions with non-zero VEVs for the $C$-moduli which could be
obtained by cancelling field-dependent FI-terms associated with anomalous $U(1)$ factors.
By the cross structure of the combined moduli space~\cite{AGLO1,AGLO2,Anderson:2011ty},
this in turn implies stabilisation of at most $h^{1,2}-h^{1,2}_{\rm hol}$ complex structure moduli.
This situation is particularly relevant for the case of heterotic orbifold compactifications which often have only
a few untwisted $Z$-moduli. In this case it seems possible to stabilise all gauge bundle moduli
and the small total number of untwisted complex structure moduli using only higher-order terms in (\ref{eq:dC00}) and a sufficient number of D-terms from anomalous $U(1)$ factors~\cite{Buchmuller:2005jr,Buchmuller:2006ik,Lebedev:2006kn,Lebedev:2007hv,Lebedev:2008un}.

In the rest of the paper we will focus on the generic situation where this stabilisation procedure fixes
all the gauge bundle moduli and some, but not all, complex structure moduli, so that
$h^{1,2}_{\rm hol}>0$ $Z$-moduli are still left flat. Furthermore, even if $h^{1,2}_{\rm hol}=0$,
it could still be that some $C$-moduli are fixed at zero VEV, implying that the complex structure moduli
could still be flat (see Fig. \ref{Plot0}).

\subsection{Corrections beyond tree-level}
\label{sec:SUSYvacNonPert}

Given that the remaining $h^{1,2}_{\rm hol}>0$ $Z$-moduli cannot be fixed at tree-level
by using quantised fluxes (since $|W_0|\neq 0$ would induce a runaway for both $s$ and $\vo$),
let us focus on perturbative and non-perturbative corrections to the scalar potential. We shall proceed in two steps,
showing first how to fix the complex structure moduli and the dilaton by the inclusion of an $S$-dependent gaugino
condensate, and then explaining how to stabilise the K\"ahler moduli by an interplay of world-sheet instantons
and threshold corrections to the gauge kinetic function.
For the time being, we shall neglect perturbative corrections to the K\"ahler potential (either $\alpha'$ or $g_s$)
since these generically break supersymmetry, and so we shall include them only in section \ref{NonSusyVacua}
where we shall study supersymmetry breaking vacua.

\subsubsection{Step 1: $Z$ and $S$ stabilisation by gaugino condensation}

Let us add a single $S$-dependent gaugino condensate to the superpotential and determine
how this term modifies the tree-level picture:
\be
W  = W_{\rm flux}+W_{\rm gc} = \int_X H\wedge\Omega + A(Z)\, e^{- \lambda S}\,.
\ee
The K\"ahler-covariant derivatives now become:
\bea
D_{Z^{\alpha}}W & = & {\rm i}\,b_\alpha(Z) +e^{- \lambda S} \left[ \partial_{\alpha}A(Z) -k_{\alpha}(Z,\bar{Z})A(Z)\right], \\
D_{S}W & = & -\frac{1}{2s}\left[{\rm i}\,a(Z)+(2  \lambda s +1)A(Z)\, e^{- \lambda S}\right], \\
D_{T_i} W & = &  -\frac{{\rm i}\,a(Z)+A(Z)\, e^{- \lambda S}}{4\vo}\,k_{ijk} t_j t_k\,.
\label{DTW}
\eea
The potential is again of the no-scale type (i.e. given by the first equality of (\ref{eq:noscale})).
At the minimum the complex structure moduli will be frozen at the solution to:
\be
D_{Z^{\alpha}} W=0\quad\Leftrightarrow\quad {\rm i}\,b_\alpha(Z)
= e^{- \lambda S} \left[k_{\alpha}(Z,\bar{Z})A(Z)-\partial_{\alpha}A(Z)\right],
\label{FixZ}
\ee
and now the dilaton is not forced anymore to run-away to infinity:
\be
D_S W=0\quad\Leftrightarrow\quad
W_0\equiv {\rm i}\, a(Z)= - (2  \lambda s +1)A(Z)\, e^{- \lambda S}\,.
\label{FixS}
\ee
The potential for the K\"ahler moduli is flat, resulting in a Minkowski vacuum
with broken supersymmetry since substituting (\ref{FixS}) into (\ref{DTW}) one finds:
\be
D_{T_i} W= -\left(\frac{2\lambda s}{2\lambda s+1}\right)\frac{W_0}{4\vo}\,k_{ijk} t_j t_k\,.
\label{Teq}
\ee
The previous expression for $W_0\neq 0$, finite volume and $t_i > 1$ $\forall i$,
gives $D_{T_i} W\neq 0$ for a generic point in moduli space.

Let us comment now on the possibility to satisfy (\ref{FixS}) at the physical point $\langle s\rangle\simeq 2$
that corresponds to $\alpha_\GUT^{-1}\simeq 25$. Setting $A=1$ and $\lambda = 8\pi^2/N$ where $N$ is the
rank of the $SU(N)$ condensing gauge group, we have (fixing the axion $a$ at $\lambda \langle a\rangle =\pi$):
\be
W_0 = \left(\frac{16 \pi^2 \langle s\rangle}{N} +1\right)\, e^{- \frac{8\pi^2}{N} \langle s\rangle} \,.
\ee
As an illustrative example, for $\langle s\rangle\simeq 2$ and $N=5$, the previous expression would give
$W_0 \simeq 10^{-12}$, which for $\vo\simeq 20$ corresponds to a gravitino mass of the order
$m_{3/2} = W_0/(\sqrt{2s\vo}) \simeq 330$ TeV. On the other hand,
for $N=30$ (as in the case of $E_8$), one would obtain $W_0 \simeq 0.06$ corresponding
to a GUT-scale gravitino mass: $m_{3/2}\simeq 10^{16}$ GeV.
Due to the absence of Ramond-Ramond fluxes, there is in general no freedom to tune the heterotic flux superpotential $W_0$
to values much smaller than unity, implying that heterotic CY compactifications
generically predict a gravitino mass close to the GUT scale.
As we already pointed out, low-energy supersymmetry could instead be obtained in the particular cases
when $h^{1,2}_{\rm hol}=0$ so that one does not need to turn on $W_0\neq 0$ to fix all the $Z$-moduli,
in orbifold constructions or in compactifications on non-complex manifolds.

A possible way to obtain fractional values of $W_0$ of order $0.1 - 0.01$
has been described in \cite{GKLM} where the authors considered a trivial $B_2$ field
and a rigid 3-cycle $\Sigma_3$ such that the integral of $H$ over $\Sigma_3$ (ignoring the contribution from the spin connection):
\be
\int_{\Sigma_3} H \simeq - \int_{\Sigma_3} {\rm CS}(A)\,,
\ee
gives rise to a fractional flux.\footnote{Note that these flux quanta are well-defined quantities
even if $H$ is not closed since a rigid homology class admits just one representative}
Stabilisation of all complex structure moduli would then require scanning the three-form flux over all cycles to search for VEVs $\langle Z_\alpha\rangle$ such that the overall $(0,3)$-contribution to the superpotential~\eqref{eq:Wflux} is of the order of the fractional Chern-Simons contribution or smaller.\footnote{For the purpose of an explicit demonstration of such vacua one may rely on CYs arising in Greene-Plesser pairs of manifolds related by mirror symmetry~\cite{Greene:1990ud,Candelas:1994hw,Candelas:2000fq}. CY mirror pairs arising from the Greene-Plesser construction have their complex structure moduli space partitioned by a typically large discrete symmetry $\Gamma$ into an invariant subspace and its complement. One can then show that the periods of the invariant subspace depend at higher-order non-trivially on all the $\Gamma$-non-invariant complex structure moduli. If the $\Gamma$-invariant subspace is of low dimensionality (as is the case e.g. of the CY $\mathbb{CP}^4_{11169}[18]$ as discussed in~\cite{Giryavets:2003vd,Denef:2004dm}), then turning on the relatively few fluxes on the invariant subspace is enough to stabilise \emph{all} complex structure moduli at an isolated minimum~\cite{Giryavets:2003vd,Denef:2004dm}. On such a CY manifold one can therefore stabilise all $Z$-moduli by just turning a few fractional Chern-Simons $(0,3)$-type fluxes on the cycles of the invariant subspace, which can serve to demonstrate the existence of such vacua.}

\subsubsection{Step 2: $T$ stabilisation by worldsheet instantons and threshold effects}
\label{SecTStab}

The K\"ahler moduli can develop a potential either by loop corrections to the gauge kinetic function
or via worldsheet instantons. Let us start considering the case with just threshold effects.
\newline

\emph{Threshold effects}: The potential generated by gaugino condensation takes the form:
\be
W_{\rm gc} = A(Z)\,e^{-\lambda \left(S-\frac{\beta_i}{2}\,T_i\right)}\,,
\ee
lifting the $T$-moduli and modifying (\ref{Teq}) into:
\be
D_{T_i}W = -\frac{\lambda\,W_0}{2(2  \lambda s +1)}\left[\beta_i
+ \frac{s}{\vo}\,k_{ijk} t_j t_k\right]=0\quad\Leftrightarrow\quad \beta_i = - \frac{s}{\vo}\,k_{ijk} t_j t_k\,.
\ee
This result, in turn, gives:
\be
{\rm Re}\left(f_{\rm hid}^{\rm 1-loop}\right) = -\frac{\beta_i}{2}\,t_i = \frac{s}{2\vo}\,k_{ijk} t_i t_j t_k
= 3\, s = 3 \,{\rm Re}\left(f_{\rm hid}^{\rm tree}\right),
\ee
implying that perturbation theory in the hidden sector is not under control since the one-loop contribution is bigger than
the tree-level one. Moreover the gauge kinetic function of the visible sector becomes negative:
\be
{\rm Re}\left(f_{\rm vis}\right) = g_{\rm vis}^{-2}= s + \frac{\beta_i}{2}\,t_i = -2 s <0\,,
\ee
meaning that the positive tree-level contribution is driven to negative values by threshold effects.
Actually, before becoming negative, $g_{\rm vis}^{-2}$ will vanish corresponding to a strong coupling transition
whose understanding is not very clear \cite{GKLM}.
Note that we neglected D-terms since, due to the relation (\ref{FDrelation}), if present,
they would also cause the same problems. Let us see now how these control issues can be addressed
by including worldsheet instantons \cite{CKL}.

\emph{Threshold effects and worldsheet instantons}: The new total non-perturbative superpotential reads:
\be
W_{\rm np} = A(Z)\,e^{-\lambda \left(S-\frac{\beta_i}{2}\,T_i\right)} + B(Z)\,e^{-\mu\,T_\ast}\,,
\ee
where we included the contribution of a single worldsheet instanton dependent on $T_\ast$.
In general, one could have more non-perturbative $\alpha'$ contributions, but we shall here
show that just one worldsheet instanton is enough to overcome the previous problems.
The new K\"ahler covariant derivatives become:
\bea\label{eq:FtermsNonPert}
D_{Z^{\alpha}}W & = & {\rm i}\,b_\alpha(Z)+W_{\rm gc} \left[ \frac{\partial_{\alpha}A(Z)}{A(Z)} - k_{\alpha}(Z,\bar{Z})\right]
+W_{\rm wi} \left[ \frac{\partial_{\alpha}B(Z)}{B(Z)} - k_{\alpha}(Z,\bar{Z})\right], \label{ZFterms} \\
D_{S}W & = & -\frac{1}{2s}\left[W_0+(2  \lambda s +1)W_{\rm gc}+W_{\rm wi}\right], \\
D_{T_p} W & = & \lambda\frac{\beta_p}{2}\,W_{\rm gc} -\frac{W_0 + W_{\rm gc} + W_{\rm wi}}{4\vo}\,k_{pjk} t_j t_k\qquad p\neq \ast\,,\\
D_{T_\ast} W & = & \lambda\frac{\beta_\ast}{2}\,W_{\rm gc}-\mu W_{\rm wi} -\frac{W_0 + W_{\rm gc} + W_{\rm wi}}{4\vo}\,k_{\ast jk} t_j t_k\,.
\eea
The solutions describing supersymmetric vacua with vanishing F-terms are:
\bea
{\rm i}\,b_\alpha(Z)&=& W_{\rm gc} \left[k_{\alpha}(Z,\bar{Z})- \frac{\partial_{\alpha}A(Z)}{A(Z)}\right]
+W_{\rm wi} \left[k_{\alpha}(Z,\bar{Z})- \frac{\partial_{\alpha}B(Z)}{B(Z)}\right],
\label{GenFixZ} \\
W_0 &=& - (2  \lambda s +1)W_{\rm gc}- W_{\rm wi}\,, \label{dilatonfix}\\
\beta_p&=&-\frac{s}{\vo}\,k_{pjk} t_j t_k\qquad p\neq \ast\,, \\
\beta_\ast &=& -\frac{s}{\vo}\,k_{\ast jk} t_j t_k +2 R\,,\quad R\equiv\frac{\mu W_{\rm wi}}{\lambda W_{\rm gc}}\,.
\eea

It is important to note that the total superpotential $W=W_0+W_{\rm gc}+W_{\rm wi}\ne 0$.
Indeed if this were zero the dilaton would not be stabilised (see (\ref{dilatonfix})).
This of course means that the supersymmetric vacua are AdS in contrast to Strominger's classical analysis \cite{Strominger:1986uh}.

The hidden and visible sector gauge kinetic functions now improve their behaviour since they look like:
\be
{\rm Re}\left(f_{\rm hid}^{\rm 1-loop}\right) = -\frac{\beta_i}{2}\,t_i =
3\,s - R\,t_\ast = 3 \,{\rm Re}\left(f_{\rm hid}^{\rm tree}\right) - R\,t_\ast\,,
\ee
and:
\be
{\rm Re}\left(f_{\rm vis}\right) = -2 s + R\,t_\ast\,.
\ee
Thus there is a regime where the hidden sector is weakly coupled
and the real part of the gauge kinetic function of the visible sector (as well as that of the hidden sector) stays positive for:
\be
2\,s\ll R\,t_\ast\ll 4\,s\,,
\ee
which points towards values $R\,t_\ast\simeq 3\,s$. In fact, in this regime,
not only ${\rm Re}\left(f_{\rm vis}\right)>0$ and ${\rm Re}\left(f_{\rm hid}\right)>0$,
but also:
\be
\left|\frac{{\rm Re}\left(f_{\rm hid}^{\rm 1-loop}\right)}{{\rm Re}\left(f_{\rm hid}^{\rm tree}\right)}\right|
= \left|\frac{{\rm Re}\left(f_{\rm vis}^{\rm 1-loop}\right)}{{\rm Re}\left(f_{\rm vis}^{\rm tree}\right)}\right|=
\left|3-\frac{R\,t_\ast}{s}\right|\ll 1\,.
\ee

\subsubsection{Tuning the Calabi-Yau condition}

As pointed out in \cite{BdA}, in the absence of worldsheet instantons and
for $\partial_{\alpha} A(Z)=0$, eq.~\eqref{eq:FtermsNonPert} reduces to:
\be
{\rm i}\,b_\alpha = W_{\rm gc} \, k_{\alpha}(Z,\bar{Z})\neq 0\,.
\ee
This induces a $(2,1)$-component of $H$ (harmonic) that should vanish according to Strominger's analysis \cite{Strominger:1986uh}.
However from (\ref{GenFixZ}), one may speculate that the CY condition can be preserved by
envisaging a situation where one tunes the flux quanta such that
$b_\alpha=0$ $\forall \alpha=1,...,h^{1,2}_{\rm hol}$ corresponding to $H^{2,1}=0$. The complex structure moduli would then be fixed by:
\be
D_{Z^\alpha} W=0\quad\Leftrightarrow\quad
\frac{W_{\rm wi}}{W_{\rm gc}}
= -\frac{1- \frac{\partial_{\alpha}A(Z)}{A(Z)k_{\alpha}(Z,\bar{Z})}}
{1- \frac{\partial_{\alpha}B(Z)}{B(Z)k_{\alpha}(Z,\bar{Z})}}\,.
\ee
However now we have $4\,h^{1,2}_{\rm hol}$ real equations determining $2\,h^{1,2}_{\rm hol}$ real complex structure moduli.
Obviously the system has no solution unless we scan over the integer fluxes. However there are only $2\,h^{1,2}+2$ integer fluxes.
Thus we have only the freedom to scan over $Q=2\left(h^{1,2}-h^{1,2}_{\rm hol}+1\right)$ integers while all $2\,h^{1,2}_{\rm hol}$ real complex structure moduli
as well as all but $Q$ of the integers (i.e. $2\,h^{1,2}_{\rm hol}$ of them) must emerge as solutions to these non-linear equations.
Thus we do not think that it is possible to have $b_\alpha=0$ in the presence of these non-perturbative terms.
However, this condition emerges only on demanding a supersymmetric solution to the classical 10D equations,
and so our 4D analysis cannot be expected to satisfy these classical conditions once non-perturbative effects are included.

\subsection{Flux vacua counting}

Let us clarify here a crucial difference between type IIB and heterotic string theory regarding complex structure
stabilisation with three-form flux. The F-term equations \eqref{ZFterms} comprise $2\,h^{1,2}$ conditions for $2\,h^{1,2}$ real variables
(setting now $h^{1,2}_{\rm hol}=h^{1,2}$ for ease of comparison with type IIB).
A non-trivial $H$-flux yields exactly $2\,h^{1,2}$ independent flux quanta (up to the two related to the overall scaling of $\Omega(Z)$) generically supplying the non-linear system of $h^{1,2}$ complex F-term conditions for the $2\,h^{1,2}$ complex structure moduli. However, the existence of a finite number of isolated solutions for such non-linear systems with as many equations as variables (rendering the system `well behaved') is not guaranteed. One expects therefore
that most of the available freedom of choice among the $2\,h^{1,2}$ $H$-fluxes is used up to find a relatively small number of isolated solutions for the complex structure moduli where all of them sit safely in the regime of large complex structure. Generically, this precludes the possibility of using the $H$-flux discretuum for tuning a very small VEV of $W_{\rm flux}$.

Note that this is different in the type IIB context. There, the availability of RR three-form flux $F_3$ supplies an \emph{additional} set of $2\,h^{2,1}$ fluxes for an overall discretuum made up from $4\,h^{1,2}$ fluxes. We have therefore an additional set of $2\,h^{1,2}$ discrete parameters available for tuning $W_{\rm flux}$ while keeping a given well-behaved complex structure moduli vacuum. Consequently, after having used $2\,h^{1,2}$ flux parameters to construct a viable complex structure vacuum, we can use the additional $2\,h^{1,2}$ flux quanta to construct a `discrete $2\,h^{1,2}$-parameter family' of complex structure vacua, which allows for exponential tuning of $W_{\rm flux}$.

Finally we note that in the heterotic case the unavailability of any additional freedom in the flux choice after fixing the $Z$-moduli,
means that we have to depend on the far more restricted choices that are available in the solution space of the complex structure moduli.
As mentioned before, one needs to scan over the $H$ flux integers in order to find $2\,h^{1,2}$ acceptable (i.e. in the geometric regime) real solutions to the $2\,h^{1,2}$ non-linear equations $D_{Z^\alpha}W=0$. The size of the solution set that we get is likely to be much smaller than the size of the original set of flux integers. Thus even if we had started with, let us say, $h^{1,2}=\mc{O}(100)$ and let each flux scan over 1 to 10, the number of acceptable fluxes are likely to be far smaller than what is required to tune the cosmological constant. It should also be emphasised here that the only source of tuning that is available after all the low-energy contributions to the vacuum energy are included, has to come from these fluxes.

\section{Supersymmetry breaking vacua}
\label{NonSusyVacua}

In this section we shall show the existence of new Minkowski vacua with spontaneous supersymmetry breaking
along the K\"ahler moduli directions. The strategy is to perform moduli stabilisation in two steps as follows:
\begin{itemize}
\item Step 1: Fix at leading order some of the moduli supersymmetrically (all the $h^{1,2}_{\rm hol}>0$
complex structure moduli, the dilaton and some K\"ahler moduli) at a high scale.

\item Step 2: Stabilise the remaining light moduli at a lower scale breaking supersymmetry mainly along the K\"ahler directions
by the inclusion of $\alpha'$ corrections to the K\"ahler potential in a way similar to type IIB.
\end{itemize}

In subsection~\ref{sec:alpha1} we shall consider the contributions to the scalar potential generated by
fluxes, non-perturbative effects and threshold corrections showing that there exist no supersymmetry breaking minimum
which lies in the regime of validity of the effective field theory. However, in subsection~\ref{sec:alpha2} we shall
describe how this situation improves by the inclusion of $\alpha'$ corrections to the K\"ahler potential
which yield trustworthy Minkowski vacua (see subsection~\ref{sec:MinkSol}) where supersymmetry is spontaneously broken
by the F-terms of the K\"ahler moduli.\footnote{See \cite{AQ} for another attempt to fix the heterotic moduli via the inclusion of $\alpha'$ effects.}
In subsection~\ref{Dtermpot} we shall explain what is the r\^ole played by D-terms in our stabilisation procedure.
Let us finally stress that this new procedure to obtain supersymmetry breaking vacua is completely orthogonal to
the way the complex structure moduli are fixed, and so our results apply also to the case with $h^{1,2}_{\rm hol}=0$
where there is no need to turn on quantised background fluxes to fix the $Z$-moduli.

\subsection{Fluxes, non-perturbative effects and threshold corrections}
\label{sec:alpha1}

In this section we shall derive the general expression for the scalar potential
including fluxes, non-perturbative effects (both gaugino condensation and world-sheet instantons)
and threshold corrections for a CY three-fold whose volume is given by:
\be
\vo = k_b t_b^3 - k_s t_s^3\,.
\label{volform}
\ee
The superpotential and the K\"ahler potential look like
(neglecting a possible $Z$-dependence of $A$ and $B$ and setting for simplicity $\beta_s=0$):
\bea
W&=& W_{\rm flux}(Z) + A\, e^{-\lambda \left(S-\frac{\beta_b}{2}\,T_b\right)}+B\,e^{-\mu\,T_s}\,, \\
K&=& -\ln\vo-\ln\left(S+\bar{S}\right)+K_{\rm cs}(Z,\bar Z)\,.
\eea
Performing the following field redefinition:
\be
\Phi \equiv S-\frac{\beta_b}{2}\,T_b\,,
\ee
$W$ and $K$ take the form:
\bea
W&=& W_{\rm flux}(Z) + A\, e^{-\lambda\,\Phi}+B\,e^{-\mu\,T_s}\,, \label{Wsimple} \\
K&=& -\ln\vo-\ln\left[\Phi+\bar{\Phi}+\frac{\beta_b}{2}\left(T_b+\bar{T}_b\right)\right]+K_{\rm cs}(Z,\bar Z)\,.
\label{Ksimple}
\eea

\subsubsection{Derivation of the F-term potential}

The F-term scalar potential turns out to be:
\bea
V&=&e^K \left[\sum_Z K^{\alpha\bar{\beta}}D_\alpha W D_{\bar{\beta}}\bar{W} + K^{\Phi\bar{\Phi}}D_\Phi W D_{\bar{\Phi}}\bar{W} \right. \nn \\
&& \left. +\left( K^{\Phi\bar{T}_b} K_{\bar{T}_b} +K^{\Phi\bar{T}_s} K_{\bar{T}_s}\right) \left(\bar{W} D_\Phi W + W D_{\bar{\Phi}}\bar{W}\right)
\right. \nn \\
&& \left. +K^{\Phi\bar{T}_s} \partial_{\bar{T}_s}\bar{W} D_\Phi W+ K^{T_s\bar{\Phi}}\partial_{T_s} W D_{\bar{\Phi}}\bar{W}\right. \nn \\
&& \left. + |W|^2 \left(\sum_T K^{i\bar{j}} K_i  K_{\bar{j}} -3\right)\right. \nn \\
&& \left. + \left(K^{T_s\bar{T}_b} K_{\bar{T}_b} +K^{T_s\bar{T}_s} K_{\bar{T}_s}\right)
\left(\bar{W}\partial_{T_s} W +W  \partial_{\bar{T}_s}\bar{W}\right) \right. \nn \\
&& \left. +K^{T_s\bar{T}_s}\partial_{T_s} W \partial_{\bar{T}_s}\bar{W}\right]. \nn
\eea
Let us consider the limit:
\be
\left|{\rm Re}\left(f_{\rm hid}^{\rm 1-loop}\right)\right|\ll {\rm Re}\left(f_{\rm hid}^{\rm tree}\right)
\quad\Leftrightarrow\quad \frac{\beta_b}{2}\,t_b\ll s\,,
\ee
which implies (defining $\Phi=\phi+{\rm i}\psi$):
\be
\epsilon_\phi \equiv \frac{\beta_b\, t_b}{2\phi} =\frac{\beta_b\,t_b}{2(s - \frac{\beta_b}{2}\,t_b)}
=-\frac{1}{1-\frac{2 s}{\beta_b\,t_b}}\simeq \frac{\beta_b\,t_b}{2 s}\ll 1\,,
\label{def}
\ee
together with:
\be
t_b\sim\mc{O}(10) > t_s\sim\mc{O}(1) \qquad\Rightarrow\qquad\epsilon_s  \equiv \frac{k_s t_s^3}{k_b t_b^3} \ll 1\,.
\label{DEF}
\ee
We can then expand the relevant terms as:
\be
K^{\Phi\bar{\Phi}} = 4 \phi^2 \left(1 + 2\epsilon_\phi + \frac{4\epsilon_\phi^2}{3}
+ \frac{\epsilon_s\epsilon_\phi^2}{6}\right),\qquad K^{\Phi\bar{T}_s}= K^{T_s\bar{\Phi}}=-2\epsilon_\phi \phi t_s\,,
\ee

\be
K^{\Phi\bar{T}_b} K_{\bar{T}_b} +K^{\Phi\bar{T}_s} K_{\bar{T}_s}=
\frac{2\epsilon_\phi \phi}{1+\epsilon_\phi}\left(1 + \frac{4\epsilon_\phi}{3}
+ \frac{\epsilon_s \epsilon_\phi}{6}\right).
\ee
The no-scale structure gets broken by loop effects:
\be
\sum_T K^{i\bar{j}} K_i  K_{\bar{j}} -3 = \frac{2\epsilon_\phi}{(1 + \epsilon_\phi)^2}
\left( 1 + \frac{7 \epsilon_\phi}{6}
+ \frac{\epsilon_\phi \epsilon_s}{12}\right).
\ee
Note that one correctly recovers the no-scale cancellation for $\beta_b=0$ $\Leftrightarrow$ $\epsilon_\phi=0$.
Other relevant terms are:
\be
K^{T_s\bar{T}_b} K_{\bar{T}_b} +K^{T_s\bar{T}_s} K_{\bar{T}_s}= -2\, t_s\left(\frac{1 + 3\epsilon_\phi/2}{1 + \epsilon_\phi}\right),
\qquad K^{T_s\bar{T}_s}=\frac{2 t_s^2}{3\epsilon_s}\left(1+2\epsilon_s\right).
\ee
We shall look for minima in the region $\vo \sim W_{\rm flux}\,e^{\,\mu \,t_s}$ implying that
$W_{\rm wi}\sim \epsilon_s\,W_{\rm flux}\ll W_{\rm flux}\sim W_{\rm gc}$. The relevant derivatives scale as:
\be
\partial_{Z^\alpha}W\sim W_{\rm flux}\,,\qquad \partial_\Phi W\sim W_{\rm gc}\sim W_{\rm flux}\,,
\qquad\partial_{T_s}W \sim W_{\rm wi} \sim \epsilon_s\,W_{\rm flux}\,.
\ee
Therefore the F-term scalar potential can be expanded in the small parameters $\epsilon_\phi$ and $\epsilon_s$ as:
\be
V = V_0 + \epsilon V_1 + \epsilon^2 V_2+...
\ee
where (defining $\hat{W}=W_{\rm flux}+W_{\rm gc}$):
\be
V_0=e^K \left(\sum_Z K^{\alpha\bar{\beta}}D_\alpha \hat{W} D_{\bar{\beta}}\bar{\hat{W}}
+4 \phi^2 D_\Phi \hat{W} D_{\bar{\Phi}}\bar{\hat{W}} \right)\sim \mc{O}\left(e^K |W_{\rm flux}|^2\right), \nn
\ee
and:
\bea
\epsilon V_1&=&e^K \left[\sum_Z K^{\alpha\bar{\beta}}\left(D_\alpha \hat{W} D_{\bar{\beta}}\bar{W}_{\rm wi}
 +D_\alpha W_{\rm wi} D_{\bar{\beta}}\bar{\hat{W}}\right)+4 \phi^2 \left(D_\Phi \hat{W} D_{\bar{\Phi}}\bar{W}_{\rm wi}+D_\Phi W_{\rm wi} D_{\bar{\Phi}}\bar{\hat{W}}\right) \right. \nn \\
&& \left. +8  \epsilon_\phi \phi^2 D_\Phi \hat{W} D_{\bar{\Phi}}\bar{\hat{W}}+ 2\epsilon_\phi \phi
\left(\bar{\hat{W}} D_\Phi \hat{W} + \hat{W} D_{\bar{\Phi}}\bar{\hat{W}}\right)
\right. \nn \\
&& \left. + 2 |\hat{W}|^2 \epsilon_\phi-2\, t_s \left(\bar{\hat{W}}\partial_{T_s} W +\hat{W}  \partial_{\bar{T}_s}\bar{W}\right)
+\frac{2 t_s^2}{3\epsilon_s} \,\partial_{T_s} W \partial_{\bar{T}_s}\bar{W}\right] \sim \mc{O}\left(\epsilon \,e^K |W_{\rm flux}|^2\right), \nn
\eea
and:
\bea
\epsilon^2 V_2&=&e^K \left[\sum_Z K^{\alpha\bar{\beta}} D_\alpha W_{\rm wi} D_{\bar{\beta}}\bar{W}_{\rm wi}
 +4 \phi^2 D_\Phi W_{\rm wi} D_{\bar{\Phi}}\bar{W}_{\rm wi} +8\epsilon_\phi \phi^2\left(D_\Phi \hat{W} D_{\bar{\Phi}}\bar{W}_{\rm wi}+h.c.\right)\right. \nn \\
&& \left. +\frac{16}{3}\,\epsilon_\phi^2 \phi^2 D_\Phi \hat{W} D_{\bar{\Phi}}\bar{\hat{W}}+ 2\epsilon_\phi \phi \left(\bar{\hat{W}} D_\Phi W_{\rm wi} + W_{\rm wi} D_{\bar{\Phi}}\bar{\hat{W}}+ h.c.\right)+\frac{2\epsilon_\phi^2\phi}{3} \left(\bar{\hat{W}} D_\Phi \hat{W} + h.c.\right)
\right. \nn \\
&& \left. -2\epsilon_\phi \phi t_s\left( \partial_{\bar{T}_s}\bar{W} D_\Phi \hat{W}+ h.c.\right) + 2\epsilon_\phi\left(\hat{W}\bar{W}_{\rm wi}+h.c.\right)
- \frac{5 \epsilon_\phi^2}{3}\,|\hat{W}|^2 \right. \nn \\
&& \left. -2\, t_s \left(\bar{W}_{\rm wi}\partial_{T_s} W +h.c.\right) -t_s \epsilon_\phi
\left(\bar{\hat{W}}\partial_{T_s} W +h.c.\right)+\frac{4 t_s^2}{3}\, \partial_{T_s} W \partial_{\bar{T}_s}\bar{W}\right]
\sim \mc{O}\left(\epsilon^2 e^K |W_{\rm flux}|^2\right). \nn
\eea

\subsubsection{Moduli stabilisation}

Let us perform moduli stabilisation in two steps.

\textit{Step 1}: We stabilise the $\Phi$ and $Z$-moduli by imposing $D_{Z^\alpha}\hat{W}=D_\Phi\hat{W}=0$
thus minimising the leading order term in the potential.
We then substitute this solution in the scalar potential obtaining $V_0=0$ whereas the other
contributions take the form:
\bea
\epsilon V_1 &=&e^K \left[2 |\hat{W}|^2 \epsilon_\phi-2\, t_s \left(\bar{\hat{W}}\partial_{T_s} W +\hat{W}  \partial_{\bar{T}_s}\bar{W}\right)
+\frac{2 t_s^2}{3\epsilon_s} \,\partial_{T_s} W \partial_{\bar{T}_s}\bar{W}\right], \nn
\eea
and:
\bea
\epsilon^2 V_2 &=&e^K \left[\sum_Z K^{\alpha\bar{\beta}} D_\alpha W_{\rm wi} D_{\bar{\beta}}\bar{W}_{\rm wi}
 +4 \phi^2 D_\Phi W_{\rm wi} D_{\bar{\Phi}}\bar{W}_{\rm wi} \right. \nn \\
&& \left. + 2\epsilon_\phi \phi \left(\bar{\hat{W}} D_\Phi W_{\rm wi}+ h.c.\right) + 2\epsilon_\phi\left(\hat{W}\bar{W}_{\rm wi}+h.c.\right)
- \frac{5 \epsilon_\phi^2}{3}\,|\hat{W}|^2 \right. \nn \\
&& \left. -2\, t_s \left(\bar{W}_{\rm wi}\partial_{T_s} W +h.c.\right) -t_s \epsilon_\phi
\left(\bar{\hat{W}}\partial_{T_s} W +h.c.\right)+\frac{4 t_s^2}{3}\, \partial_{T_s} W \partial_{\bar{T}_s}\bar{W}\right]. \nn
\eea

\textit{Step 2}: We stabilise the $T$-moduli at order $\mc{O}(\epsilon)$ breaking supersymmetry.
Writing $T_s= t_s +{\rm i} a_s$, $W_0^{\rm eff}= |W_0^{\rm eff}|e^{{\rm i} \theta_W}$ and $B= |B|e^{{\rm i}\theta_B}$,
and setting $e^{\langle K_{\rm cs} \rangle}=1$, the explicit form of the scalar potential
at $\mc{O}(\epsilon)$ is:
\be
V = \left[ \frac{A_1}{\vo^{2/3}}
+\frac{|A_2|}{|W_0^{\rm eff}|}  \cos (\theta_B-\theta_W-\mu a_s) \frac{t_s\, e^{-\mu t_s}}{\vo}
+\frac{|A_3|}{|W_0^{\rm eff}|^2} \frac{e^{-2 \mu t_s}}{t_s}\right]\frac{|W_0^{\rm eff}|^2}{\langle\phi\rangle}\,,   \label{VOe}
\ee
with:
\be
A_1 \equiv \frac{\beta_b}{2\langle\phi\rangle k_b^{1/3}},\qquad |A_2|\equiv 2 \,\mu |B|,
\qquad |A_3|\equiv \frac{\mu^2 |B|^2}{3 k_s}\,, \label{Param}
\ee
where we have defined $W_0^{\rm eff}\equiv \langle \hat{W}\rangle = \langle W_{\rm flux} + W_{\rm gc}\rangle$.
The axion $a_s$ is minimised at $\mu\langle a_s\rangle = \theta_B-\theta_W-\pi$ so that (\ref{VOe}) reduces to:
\be
V =  \left[\frac{A_1}{\vo^{2/3}}
- \frac{|A_2|}{|W_0^{\rm eff}|}  \frac{t_s \,e^{-\mu t_s}}{\vo}
+\frac{|A_3|}{|W_0^{\rm eff}|^2} \frac{e^{-2 \mu t_s}}{t_s}\right]\frac{|W_0^{\rm eff}|^2}{\langle\phi\rangle}\,. \label{Vfin}
\ee
Minimising with respect to $t_s$ one finds:
\be
\vo  = \frac{|A_2||W_0^{\rm eff}|}{|A_3|} \frac{(\mu t_s-1)}{(2 \mu t_s+1)} \,t_s^2\, e^{\mu t_s}
\underset{\mu t_s \gg 1}{\simeq} \frac{|A_2||W_0^{\rm eff}|}{2|A_3|}\,t_s^2\, e^{\mu t_s}
= \frac{3 k_s t_s^2}{\mu |B|}\,|W_0^{\rm eff}|\, e^{\mu t_s}\,,
\label{vomin}
\ee
which implies:
\be
\mu t_s = \ln \left(\frac{\vo}{|\lambda_0|}\right)-2\ln t_s
\underset{t_s \sim \mc{O}(1)}{\simeq}\ln \left(\frac{\vo}{|\lambda_0|}\right)\equiv x(\vo)\,\qquad\text{with}\qquad
|\lambda_0|\equiv \frac{3 k_s |W_0^{\rm eff}|}{\mu |B|}\,. \label{vomin2}
\ee
Note that we can trust our effective field theory when $t_s\geq 1$,
that is when $x(\vo)\geq \mu = 2 \pi$.
Substituting (\ref{vomin}) and (\ref{vomin2}) in (\ref{Vfin}), we end up with:
\be
V =  \left[A_1\,\vo^{4/3} - |C_0|\, x(\vo)^3\right]
\frac{|W_0^{\rm eff}|^2}{\langle\phi\rangle\,\vo^2}\,,\qquad\text{where}
\qquad|C_0|\equiv \frac{3 k_s}{\mu^3}\,. \label{Vtb}
\ee
The extrema of $V$ are located at:
\be
\frac{\partial V}{\partial \vo}=0 \qquad \Leftrightarrow\qquad
A_1 \vo^{4/3} = 3 |C_0|\,x(\vo)^2 \left[x(\vo)-\frac 32\right],
\label{Vtbb}
\ee
showing that $A_1$ has to be positive, i.e. $\beta_b>0$, if we want to have a minimum at large volume,
i.e. $x(\vo)\geq 2 \pi$.
Evaluating the second derivative at these points one finds:
\be
\frac{\partial^2 V}{\partial \vo^2}>0 \qquad \Leftrightarrow\qquad 4 x^2 -15 x + 9<0\,.
\label{Vtbsec}
\ee
Hence the scalar potential has a minimum only for:
\be
\frac 34 < x(\vo) < 3\,,
\ee
provided one can find values of $\lambda_0$ that satisfy (\ref{Vtbb}) for this range of values for $\vo$.
However these minima are not trustworthy since the blow-up mode $t_s$ is fixed below
the string scale as $\langle t_s\rangle \simeq x(\vo)/(2\pi) < 3/(2\pi)$.
Moreover, the above derivation assumed a regime $x(\vo)> 2\pi$ but leads to a condition $x(\vo)< 3$ for a minimum to exist,
demonstrating the absence of a minimum for the $T$-moduli in a controlled region of the scalar potential.
This is consistent with a numerical analysis of the scalar potential~\eqref{Vfin} which shows that
in the range $3/4<x<3$ the only critical point is a saddle point with one tachyonic direction.

\subsection{Inclusion of $\alpha'$ effects}
\label{sec:alpha2}

Let us now try to improve this situation taking into account also $\alpha'$ corrections to the K\"ahler potential described in section \ref{AlphaPrime}.
Including both $\mc{O}(\alpha'^2)$ and $\mc{O}(\alpha'^3)$ effects, the K\"ahler potential for the $T$-moduli receives the following corrections:
\be
K \simeq -\ln\vo + \frac{|c_b|}{\vo^{2/3}} - \frac{\gamma_s t_s+\xi/2}{\vo}\,,\qquad\text{with}\qquad
\gamma_s\equiv |c_b|\,k_s^{1/3}-|\kappa| k_b^{1/3}\,,
\label{gsconstr}
\ee
where we have used eq. (\ref{cs}) for the expression for $c_s$.
These higher-derivative corrections break the no-scale structure as (neglecting threshold effects):
\be
\sum_T K^{i\bar{j}} K_i  K_{\bar{j}} -3 \simeq -\frac{2 |c_b|}{\vo^{2/3}}
+\frac{2 \gamma_s t_s+ 3\xi}{\vo}\,.
\ee
The scalar potential (\ref{Vfin}) gets modified and reads:
\be
V =  \left[\frac{A_1}{\vo^{2/3}}-\frac{|c_b|}{\vo^{5/3}}
- \frac{|A_2|}{|W_0^{\rm eff}|} \frac{t_s }{\vo}\,e^{-\mu t_s}
+\frac{|A_3|}{|W_0^{\rm eff}|^2} \frac{e^{-2 \mu t_s}}{t_s}
+\frac{\gamma_s t_s+3\xi/2}{\vo^2}\right]\frac{|W_0^{\rm eff}|^2}{\langle\phi\rangle}\,. \label{Vtotal}
\ee
Minimising with respect to $t_s$ we find:
\bea
\vo &=& \left(1 \pm \sqrt{
   1 + \frac{4 |A_3| \gamma_s (2 \mu t_s+1)}{|A_2|^2 t_s^2 (\mu t_s-1)^2}}\right)
   \frac{|A_2| |W_0^{\rm eff}| t_s^2 (\mu t_s -1)}{2 |A_3| (2 \mu t_s + 1)}\,e^{\mu t_s} \nn \\
&\underset{\mu t_s \gg 1}{\simeq}&
      \frac{|A_2||W_0^{\rm eff}|}{4 |A_3|}\,\left(1 + \sqrt{
   1 + \frac{c}{t_s^3}}\,\right)\,t_s^2\,e^{\mu t_s}\qquad\text{with}\qquad c=\frac{8 |A_3| \gamma_s}{|A_2|^2 \mu}=\frac{2\gamma_s}{3k_s\mu}\,,
\label{vomin3}
\eea
where we focused only on the solution which for $\gamma_s =0$ correctly reduces to (\ref{vomin}) since the
other solution can be shown to give rise to a maximum along the $t_s$ direction. Note that we did not
take an expansion for small $c/t_s^3$ even if this quantity is suppressed by $\mu t_s\gg 1$ since a large denominator
might be compensated by a large value of the unknown coefficient $\gamma_s$.
Performing the following approximation:
\be
\mu t_s = \ln\left(\frac{\vo}{|\lambda|}\right)-2\ln t_s
\underset{t_s \sim \mc{O}(1)}{\simeq}\ln\left(\frac{\vo}{|\lambda|}\right)\equiv x(\vo)\,,
\label{x}\ee
with:
\be
|\lambda|\equiv \frac{|\lambda_0|}{2}\,\left(1 + \sqrt{ 1 + \frac{c}{t_s^3}}\,\right), \nn
\ee
and substituting (\ref{vomin3}) in (\ref{Vtotal}) we end up with (in the regime $x(\vo)\gg 1$):
\be
V \simeq  \left[A_1\,\vo^{4/3}- |c_b|\,\vo^{1/3}(1-\delta\,x)
- |C|\,x^3 +\frac{3 \xi}{2}\right]\frac{|W_0^{\rm eff}|^2}{\langle\phi\rangle\vo^2}\,,
\label{Valpha}
\ee
where we have defined:
\be
|C|\equiv \frac{ |C_0|}{2}\,\left(1 + \sqrt{1 + \frac{c \mu^3}{x^3}}\right)\,,
\qquad\text{and}\qquad \delta \equiv  \frac{\gamma_s}{\mu\,|c_b|\,\vo^{1/3}}\,.
\label{Def}
\ee
Note that if we switch off the $\alpha'$ corrections by setting $|c_b|=c=\xi=0$,
the scalar potential (\ref{Valpha}) correctly reduces to (\ref{Vtb}) since $|\lambda|\to |\lambda_0|$
and $|C|\to|C_0|$.

Before trying to minimise this scalar potential, let us show two important facts:
\begin{itemize}
\item The quantity $\delta\, x$ is always smaller than unity since from (\ref{gsconstr}) one finds that:
\be
\gamma_s \leq |c_b| k_s^{1/3}\qquad\Rightarrow\qquad \delta\, x\leq \frac{k_s^{1/3} t_s}{\vo^{1/3}}
\simeq \epsilon_s^{1/3} \ll 1\,.
\ee
Therefore the term in (\ref{Valpha}) proportional to $|c_b|$ has always a positive sign.

\item If the condition $|c|/t_s^3 \ll 1$ is not satisfied, there is no minimum for realistic values of the underlying parameters.
In fact, in this case the term proportional to $|C|$ is always sub-leading with respect to the term proportional to $c_b$ since for $c\geq 0$:
\be
R \equiv \frac{|C|\,x^3}{|c_b|\,\vo^{1/3}}\leq \left(1+\sqrt{1+\frac{c}{t_s^3}}\right)\frac{t_s^3}{c}\frac{\epsilon_s^{1/3}}{x}
\ll \left(1+\sqrt{1+\frac{c}{t_s^3}}\right)\frac{t_s^3}{c}\,,
\ee
which for $c/t_s^3 \sim \mc{O}(1)$ reduces to $R \ll \mc{O}(1)$, whereas for $c/t_s^3 \gg 1$
reduces to $R \ll \sqrt{t_s^3/c}\ll 1$. On the other hand, for $c<0$, one has $|c|/t_s^3 \leq 1$ but
if $|c|/t_s^3 \sim \mc{O}(1)$, the ratio $R$ can be shown to reduce again to $R< \epsilon_s^{1/3}/x\ll 1$.
Therefore in this case the leading order scalar potential is given by (neglecting the term proportional to $\delta$):
\be
V \simeq  \left[A_1\,\vo^{4/3}- |c_b|\,\vo^{1/3}+\frac{3 \xi}{2}\right]\frac{|W_0^{\rm eff}|^2}{\langle\phi\rangle\vo^2}\,,
\label{Vnogood}
\ee
with:
\be
|c|\gtrsim t_s^3 \qquad\Rightarrow\qquad |c_b|\geq \frac{\gamma_s}{k_s^{1/3}} = \frac{3 k_s^{2/3} \mu}{2}\,c \gtrsim \frac{3 k_s^{2/3} \mu}{2}\,t_s^3\,.
\ee
However the potential (\ref{Vnogood}) has a minimum only if:
\be
\xi > \frac{5}{12}\,|c_b|\vo^{1/3}\gtrsim \frac{5}{8}\,k_s t_s^3\frac{x}{\epsilon_s^{1/3}}\gg 1\,,
\ee
which is never the case for ordinary CY three-folds with $\xi\sim \mc{O}(1)$.
As an illustrative example, for $\vo=20$, $t_s=1.5$ and $k_s= n/6$ with $n \in \mathbb{N}$, one finds $\xi \gtrsim 11\, n^{2/3} \geq 11$,
corresponding to a CY with Euler number negative and very large in absolute value: $|\chi|=2(h^{1,2}-h^{1,1})\gtrsim 4548$,
while most of the known CY manifolds have $|\chi|\lesssim \mc{O}(1000)$.
\end{itemize}

Hence we have shown that in order to have a trustable minimum we need to be in a region where $|c|\ll t_s^3$.
In this case, the scalar potential (\ref{Valpha}) simplifies to:
\be
V \simeq  \left[A_1\,\vo^{4/3}- |c_b|\,\vo^{1/3}(1-\delta\,x)- |C_0|\,x^3
+\frac{3\xi}{2}\right]\frac{|W_0^{\rm eff}|^2}{\langle\phi\rangle\vo^2}\,,
\label{Valpha2}
\ee
where we have approximated $|C|\simeq |C_0|$.
Note that the sign of the numerical coefficient $A_1$ is a priori undefined
and depends on the sign of the underlying parameter $\beta_b$.

The new extrema of $V$ are located at:
\be
A_1 \vo^{4/3} = 3 |C_0|\,x^2 \left(x-\frac 32\right) +\frac{5 |c_b|}{2}\, \vo^{1/3}\left(1- \frac{6\delta\,x}{5} + \frac{3\delta}{5} \right)- \frac{9 \xi}{2}\,,
\label{Vtbnew}
\ee
and the second derivative at these points is positive if:
\be
u(x)\equiv 12\xi-5 |c_b| \, \vo^{1/3} \left( 1- \frac{8\delta\,x}{5} + 2\delta \right)- 2 |C_0| x (4 x^2-15 x +9)>0\,.
\label{Vsecond}
\ee
Note that for $|c_b|=\delta=\xi=0$ (\ref{Vtbnew}) and (\ref{Vsecond}) correctly reduce to (\ref{Vtbb}) and (\ref{Vtbsec}) respectively.
However we shall now show that by including $\alpha'$ corrections we can find a vacuum with $x\gg 1$ where we can trust the effective field theory.

The value of the vacuum energy is:
\be
\langle V \rangle =  \frac{|W_0^{\rm eff}|^2}{2 \langle\phi\rangle\vo^2}\,v(x)\,,
\ee
where:
\be
v(x)\equiv- 6 \xi  + |C_0| x^2 \left(4 x- 9\right)+ 3 |c_b| \vo^{1/3}\left(1- \frac{4\delta\,x}{3} + \delta \right).
\ee
Let us perform the following tuning to get a Minkowski vacuum:
\be
v(x)=0\qquad\text{if}\qquad 6 \xi = 3 |c_b| \, \vo^{1/3}\left(1- \frac{4\delta\,x}{3} + \delta \right)+ |C_0| x^2 \left(4 x- 9\right),
\label{tuning}
\ee
and substitute it in (\ref{Vsecond}) obtaining:
\be
u(x)\equiv |c_b| \vo^{1/3} \left(1- 4\delta\right) + 12 |C_0| x \left(x-\frac 32\right)>0\,,
\ee
which is automatically satisfied for $\delta\ll 1$ and $x\gg 1$.
Substituting (\ref{tuning}) also in the vanishing of the first derivative (\ref{Vtbnew}), this simplifies to:
\be
4 A_1\, \vo^{4/3} = |c_b| \vo^{1/3} \left(1- 3\delta\right) + 9 |C_0| x^2\,,
\label{Min}
\ee
showing that if we want to have a Minkowski minimum $A_1$ has to be positive, i.e. $\beta_b>0$.

\subsection{Minkowski solutions}
\label{sec:MinkSol}

Let us first define our use of the term `Minkowski solutions'.
Owing to the lack of tuning freedom in the heterotic three-form flux superpotential,
achieving vacua with exponentially small vacuum energy is a real challenge. Thus we shall
use the terminology `Minkowski vacuum' to refer to
a vacuum with a cosmological constant suppressed by at least a 1-loop factor $1/(8\pi^2)\simeq 0.01$
compared to the height of the barrier in the scalar potential (of order $m_{3/2}^2M_P^2$) which protects the $T$-moduli from run-away.

\begin{figure}[ht]
\begin{center}
\epsfig{file=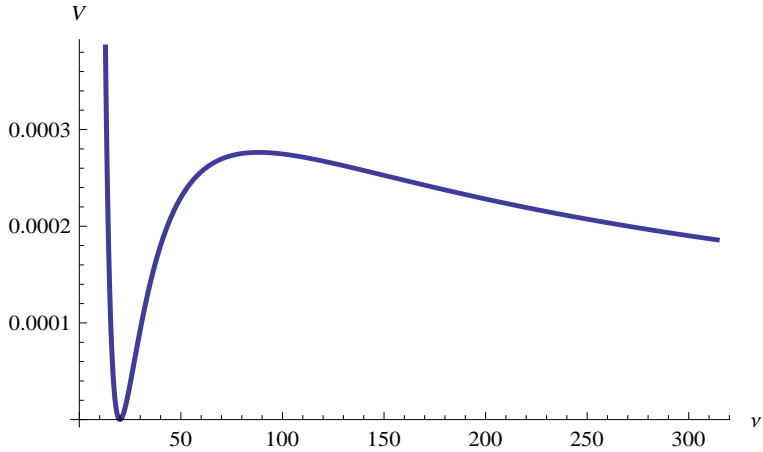, height=60mm,width=100mm}
\caption{$V$ versus $\vo$ assuming the parameters listed in the text
which give rise to a near-Minkowski vacuum with $\langle \vo \rangle = 20$
and a cosmological constant of small magnitude compared to height of the barrier set by $m_{3/2}^2M_P^2$.}
\label{Plot1}
\end{center}
\end{figure}

The solutions depend on seven underlying parameters: $k_s, k_b, \beta_b, |B|, |c_b|, |\kappa|$ and $\xi$.
We do not consider $|W_0^{\rm eff}|$ as a free variable at this stage since we fix its value at
$|W_0^{\rm eff}|=0.06$ by the phenomenological requirement of obtaining
the right GUT coupling corresponding to $\langle s \rangle \simeq \langle \phi\rangle \simeq 2$.
Let us now describe a strategy to find the values of these underlying parameters which give Minkowski vacua
for desired values of the moduli and within the regime of validity of all our approximations.
\begin{enumerate}
\item Choose the desired values for $\vo$ and $t_s$ (so fixing the value of $x=2\pi\,t_s$).
Then work out the value of $|B|$ as a function of $k_s$ from (\ref{vomin2}).

\item Choose the desired value of $t_b$ and work out the value of $k_b$ as a function of $k_s$ from (\ref{volform}).

\item Determine $|c_b|$ as a function of $k_s, \xi$ and $|\kappa|$ from (\ref{tuning}).

\item Derive the value of $\beta_b$ as a function of $k_s, \xi$ and $|\kappa|$ from (\ref{Min}).

\item Choose the values of $k_s, \xi$ and $|\kappa|$ so that all our approximations are under control, i.e.
$\epsilon_\phi$ defined in (\ref{def}) satisfies $\epsilon_\phi \ll 1$,
$\epsilon_s$ defined in (\ref{DEF}) gives $\epsilon_s \ll 1$,
$\delta$ defined in (\ref{Def}) satisfies $\delta\ll 1$ and $\epsilon_{\alpha'}\equiv \xi/(2\vo)\ll 1$.
These values of $k_s, \xi$ and $|\kappa|$ then give the values of $k_b, |B|,|c_b|$ and $\beta_b$ knowing that
this Minkowski vacuum is fully consistent.
\end{enumerate}
As an illustrative example, following this procedure we found a Minkowski vacuum (see Fig. \ref{Plot1}) located at:
\be
\langle\phi\rangle\simeq\langle s \rangle =2\,,\qquad \langle \vo \rangle = 20\,, \qquad \langle t_b\rangle \simeq 5\,,\qquad\langle t_s\rangle = 1.5\,,
\ee
for the following choice of the microscopic parameters:
\bea
k_b &=& k_s=1/6\,,\quad \beta_b\simeq 0.035\,, \quad |W_0^{\rm eff}|=0.06\,, \quad c_b=0.75\,,\quad c_s=-0.75\,, \nn \\
B&\simeq& 3\,, \quad \xi\simeq 1.49\,, \quad \mu=2\pi\quad\Rightarrow \quad\gamma_s\simeq 0.41\,,\quad \kappa=0\,.
\eea

Note that one can get dS or AdS solutions by varying $\beta_b$ either above or below its benchmark value.
Moreover, our approximations are under control since:
\be
\epsilon_\phi \simeq 0.043\,, \qquad \epsilon_s \simeq 0.027\,,\qquad \epsilon_{\alpha'}\simeq 0.037\,,\qquad \delta\simeq 0.032\,.
\ee
We stress that at the minimum these four quantities are all of the same size: $\epsilon_\phi \simeq \epsilon_s\simeq \epsilon_{\alpha'}\simeq \delta$.
This has to be the case since they weight the relative strengths of loops, non-perturbative and higher derivative effects
which all compete to give a minimum.

Moreover, we point out that there seem to be problems with the $\alpha'$ expansion since we managed to obtain a minimum
by tuning the underlying parameters in order to have the $\mc{O}(\alpha'^2)$ term
of the same order of magnitude of the $\mc{O}(\alpha'^3)$ term, and so higher order $\alpha'$ corrections
might not be negligible.

However this might not be a problem if at least one of the following is valid:
\begin{itemize}
\item The $\mc{O}(\alpha'^2)$ corrections could be eliminated by a proper redefinition of the moduli.

\item The coefficients of higher order $\alpha'$ corrections are not tuned larger than unity,
resulting in a $\alpha'$ expansion which is under control. In fact, the $\alpha'$ expansion parameter
is of order $q \vo^{-1/3}$ with $q$ an unknown coefficient. Thus $\mc{O}(\alpha'^4)$
contributions to the scalar potential can be estimated as:
\be
\frac{V_{\alpha'^4}}{V_{\alpha'^3}} \simeq \frac{2 q}{3 \xi \vo^{1/3}}\simeq 0.16\qquad\text{for}\qquad q=1\,.
\ee
\end{itemize}

\subsection{D-term potential}
\label{Dtermpot}

So far only F-terms have been taken into account.
This could be consistent since moduli-dependent D-terms might not be present in the absence of anomalous $U(1)$s,
or they might be cancelled by giving suitable VEVs to charged matter fields.

However let us see how D-terms might change the previous picture in the presence of anomalous $U(1)$s
but without introducing charged matter fields. Because of the $U(1)$-invariance of the superpotential (\ref{Wsimple}),
both $\Phi$ and $T_s$ have to be neutral. Therefore the only field which can be charged under an anomalous $U(1)$ is
$T_b$ with $q_{T_b}=4 \,c_1^b(\mc{L})\neq 0$. From (\ref{FDrelation}), this induces an FI-term of the form:
\be
\xi = -\,q_{T_b} K_b = q_{T_b} \left(\frac{k_b}{\vo}\right)^{1/3}\,,
\ee
which gives the following D-term potential:
\be
V_D = \frac{\xi^2}{{\rm Re}(f)}\simeq \frac{p}{\vo^{2/3}}
\qquad\text{with}\qquad p\equiv \frac{q_{T_b}^2\,k_b^{2/3}}{\phi}\,.
\ee
This term has the same volume scaling as the first term in (\ref{Valpha2})
which is the contribution coming from threshold effects. However the ratio between these two terms
scales as:
\be
\frac{V_{\rm threshold}}{V_D} = \frac{\epsilon_\phi}{\vo^{1/3}}\,\frac{|W_0^{\rm eff}|^2}{q_{T_b}^2\,k_b^{2/3}}\ll 1\,,
\ee
for $\epsilon_\phi \ll 1$ and $q_{T_b}\sim \mc{O}(1)$.
As an illustrative example, our explicit parameter choice would give $V_{\rm threshold}/V_D \simeq 2\cdot 10^{-4}\,q_{T_b}^{-2}$,
showing that $V_D$ is always dominant with respect to the F-term potential (\ref{Valpha2}).
In this case, $V_D$ would give a run-away for the volume direction and destroy our moduli stabilisation scenario.

As we have already pointed out, this might not be the case if there are no anomalous $U(1)$s or
if the FI-term is cancelled by a matter field VEV. There is however another way-out to this D-term problem which relies on the
possibility to fix all the moduli charged under anomalous $U(1)$s in a completely supersymmetric way,
so ensuring the vanishing of the D-term potential.
This requires $q_{T_b}=0$ and the addition of a third K\"ahler modulus $T_c$ which is charged under
an anomalous $U(1)$: $q_{T_c}\neq 0$. Let us describe this situation in the next subsection.

\subsubsection{D + F-term stabilisation}
\label{FplusD}

The K\"ahler and superpotential now read:
\bea
W&=& W_{\rm flux}(Z) + A\, e^{-\lambda\left(\Phi-\frac{\beta_c}{2}\,T_c\right)}+B\,e^{-\mu\,T_s}\,, \label{WDF} \\
K&=& -\ln\tilde{\vo}-\ln\left[\Phi+\bar{\Phi}+\frac{\beta_b}{2}\left(T_b+\bar{T}_b\right)\right]+K_{\rm cs}(Z)\,, \label{KDF}
\eea
with
\be
\tilde{\vo}=\vo - k_c t_c^3 = k_b t_b^3 - k_s t_s^3 - k_c t_c^3\,.
\ee
Note that now $\Phi$ has to get charged under an anomalous $U(1)$ so that
the hidden sector gauge kinetic function $f_{\rm hid} = \Phi-\frac{\beta_c}{2}\,T_c$ becomes gauge invariant.
In particular we will have $q_\Phi=\frac{\beta_c}{2}\,q_{T_c}$. From (\ref{FDrelation}), the FI-term looks like:
\be
\xi = -q_\Phi\,\frac{D_\Phi W}{W} -q_{T_c}\,\frac{D_{T_c} W}{W}\,,
\ee
implying that $V_D=0$ if both $\Phi$ and $T_c$ are fixed supersymmetrically.
However we have already seen that if all the K\"ahler moduli are fixed supersymmetrically via
threshold effects, then perturbation theory breaks down in the hidden sector and
the visible sector gauge kinetic function becomes negative. A way-out proposed in section \ref{SecTStab}
was to include worldsheet instantons but, given that we want to
break supersymmetry at leading order along $T_b$ and $T_s$, in order to follow this possibility we should
include a fourth modulus with worldsheet instantons. Thus this case does not look very appealing since
it requires at least four moduli.

A simpler solution can be found by noticing that the problems with ${\rm Re}\left(f_{\rm hid}^{\rm 1-loop}\right) > \,{\rm Re}\left(f_{\rm hid}^{\rm tree}\right)$ and ${\rm Re}\left(f_{\rm vis}\right)<0$ could be avoided if only some but not all of the K\"ahler moduli are
fixed supersymmetrically by threshold effects. We shall now prove that this is indeed the case if the $T$-moduli fixed in this way
are blow-up modes like $t_c$. In fact, the solution to $D_{T_c} W=0$ gives:
\be
\beta_c = - \frac{s}{\vo}\,k_{cjk} t_j t_k = - \frac{6\,s}{\vo}\,k_c t_c^2\,.
\ee
This result, in turn, gives hidden and visible sector gauge kinetic functions of the form:
\be
\frac{{\rm Re}\left(f_{\rm hid}^{\rm 1-loop}\right)}{{\rm Re}\left(f_{\rm hid}^{\rm tree}\right)}
= -\frac{\beta_b t_b}{2s} - \frac{\beta_c t_c}{2s}
= -\epsilon_\phi + 3\,\frac{k_c\,t_c^3}{\vo}
= -\epsilon_\phi + 3 \epsilon_s \ll 1\,, \nn
\ee
and:
\be
{\rm Re}\left(f_{\rm vis}\right) =  s \left( 1  + \frac{\beta_b t_b}{2 s} + \frac{\beta_c t_c}{2 c}\right)
=  s \left( 1  + \epsilon_\phi - 3\,\frac{k_c t_c^3}{\vo} \right)
=  s \left( 1  + \epsilon_\phi - 3\,\epsilon_s \right)\simeq s >0\,. \nn
\ee

\section{Moduli mass spectrum, supersymmetry breaking and soft terms}
\label{sec:Softmass}

Expanding the effective field theory around the vacua found in the previous section,
we can derive the moduli mass spectrum which turns out to be (see (\ref{x}) and (\ref{Def}) for the definitions of $x$ and $\delta$):
\bea
m_{t_s}&\simeq& m_{a_s} \simeq m_{3/2}\,x\,, \nn \\
m_{Z^\alpha}&\simeq& m_\Phi \simeq m_{3/2}\,, \nn \\
m_{t_b}&\simeq& m_{3/2}\,\delta\,, \nn \\
m_{a_b}&\simeq& 0\,.
\eea
Note that in the absence of $T_b$-dependent worldsheet instantons which would give $a_b$
a mass of order $m_{a_b}\simeq M_P\,e^{-\mu\,t_b}\simeq 10$ TeV for $t_b\simeq 5$,
this axion might be a good QCD axion candidate since it could remain a flat direction until
standard QCD non-perturbative effects give it a tiny mass.

Moreover, the stabilisation procedure described in the previous sections leads to vacua which break
supersymmetry spontaneously mainly along the K\"ahler moduli directions. In fact,
from the general expression of the F-terms and the gravitino mass:
\be
F^i = e^{K/2}K^{i\bar{j}}D_{\bar{j}} \bar{W}\qquad\text{and}\qquad
m_{3/2}= e^{K/2}|W|\simeq \frac{|W_0^{\rm eff}|}{\vo^{1/2}}\,,
\ee
we find that the K\"ahler moduli F-terms read:
\be
\frac{F^{T_b}}{t_b} = - 2\,m_{3/2}\qquad\text{and}\qquad
\frac{F^{T_s}}{t_s} \simeq \frac{m_{3/2}}{x}\,.
\ee
On the other hand, the dilaton and the complex structure moduli
are fixed supersymmetrically at leading order. However, due to the fact
that the prefactor of worldsheet instantons and $\alpha'$ effects are expected
to depend on these moduli, they would also break supersymmetry at sub-leading order
developing F-terms whose magnitude can be estimated as:\footnote{Assuming that there are no
cancellations from shifts of the minimum due to sub-leading corrections.}
\be
D_{Z^\alpha,\Phi} W \simeq D_{Z^{\alpha},\Phi} W_{\rm wi} \simeq \delta\, |W_0^{\rm eff}|\qquad\Rightarrow\qquad
F^{Z^{\alpha},\Phi} \simeq \delta\, m_{3/2}\,.
\ee
Thus we can see that supersymmetry is mainly broken along the $t_b$-direction since:
\be
\frac{F^{T_b}}{m_{3/2}}\simeq t_b \gg \frac{F^{T_s}}{m_{3/2}} \simeq \frac{t_s}{x} \gg \frac{F^{Z^\alpha,\Phi}}{m_{3/2}} \simeq \delta\,.
\ee
The goldstino is therefore mainly the $T_b$-modulino which is eaten up by the gravitino in the super-Higgs mechanism.

Soft supersymmetry breaking terms are generated in the visible sector via tree-level gravitational
interactions due to moduli mediation. Let us now derive their expressions:
\begin{itemize}
\item \textbf{Gaugino masses}: Their canonically normalised expression is given by:
\be
M_{1/2} = \frac{1}{2 {\rm Re}(f_{\rm vis})}\,F^i\partial_i f_{\rm vis}
\simeq \frac{F^\Phi}{2 \phi} + \delta\, \frac{F^{T_b}}{t_b}
\simeq \delta\, m_{3/2}\,,
\ee
showing that the gaugino masses are suppressed with respect to the gravitino mass by a factor of order
$\delta \simeq 0.03$.

\item \textbf{Scalar masses}: The canonically normalised scalar masses generated by gravity mediation read:
\be
m_{0,\,\alpha}^2 = m_{3/2}^2 - F^i \bar{F}^{\bar{j}}\partial_i \partial_{\bar{j}}\ln \tilde{K}_{\alpha}\,,
\ee
where $\tilde{K}_{\alpha}$ is the K\"ahler metric for matter fields which we assumed to be diagonal.
$\tilde{K}_{\alpha}$ is generically a function of all the moduli but we shall neglect its dependence on the dilaton
and the complex structure moduli since they give only a sub-leading contribution to supersymmetry breaking.
Hence we shall consider a K\"ahler metric for matter fields of the form $\tilde{K}_{\alpha}\simeq t_s^{-n_s}\,t_b^{-n_b}$,
where $n_s$ and $n_b$ are the so-called modular weights. In the type IIB set-up, it is possible
to determine the value of $n_b$ by requiring physical Yukawa couplings which do not depend on the large cycle
due to the localisation of the visible sector on one of the small cycles \cite{Kmatter}.
However, in the heterotic framework the situation is different. For instance, in CY compactifications close to the orbifold point the visible sector typically is constructed from split multiplets which partially live in the bulk and partially arise as twisted sector states localised at orbifold fixed points. The value of the modular weights for the different matter fields is then determined by the requirements of modular invariance.
Hence, they cannot be constrained by using an argument similar to the one in \cite{Kmatter}. We shall therefore leave them
as undetermined parameters. The scalar masses turn out to be:
\be
m_0^2 = m_{3/2}^2 \left(1-n_b -\frac{n_s}{4 x^2}\right)\,,
\ee
showing that for $x\gg 1$, the modular weight $n_b$ has to be $n_b\leq 1$ in order to avoid tachyonic squarks and sleptons.
If $n_b=1$, one has a leading order cancellation in the scalar masses which therefore get generated by the F-terms of the
small cycle $t_s$ even if $F^{T_s}\ll F^{T_b}$ (in this case one would need $n_s<0$).
This is indeed the case in type IIB models because of the no-scale structure \cite{IIBSoftTerms}.
Given that the no-scale cancellation holds in the heterotic case as well, we expect a similar cancellation to occur
in our case, i.e. $n_b=1$, with possibly the exception of twisted matter fields at orbifold fixed points, i.e. $n_b<1$ for twisted states.

\item \textbf{A-terms}: The canonically normalised A-terms look like:
\be
A_{\alpha\beta\gamma} = F^i \left[K_i + \partial_i \ln Y_{\alpha\beta\gamma}
-\partial_i \ln \left(\tilde{K}_\alpha\tilde{K}_\beta\tilde{K}_\gamma\right)\right]\,,
\ee
where $Y_{\alpha\beta\gamma}$ are the canonically unnormalised Yukawa couplings which can in principle depend on all the moduli.
Similarly to the K\"ahler metric for matter fields, we introduce two modular weights, $p_b$ and $p_s$, and we write the Yukawa couplings as
$Y_{\alpha\beta\gamma}\simeq t_b^{-p_b}\,t_s^{-p_s}$. Thus the A-terms take the form:
\be
A_{\alpha\beta\gamma} = 3 m_{3/2}\left(1+p_b-n_b+\frac{p_s}{2x}-\frac{n_s}{2x}-\frac{\delta}{3 x}\right)\,.
\ee
In the type IIB case, there is again a leading order cancellation (since $n_b=1$ and $p_b=0$ given that the Yukawa couplings
do not depend on the K\"ahler moduli due to the axionic shift-symmetry and the holomorphicity of $W$) which is again
due to the no-scale structure \cite{IIBSoftTerms}. Similarly to the scalar masses,
we expect this leading order cancellation also in the heterotic case for matter fields living in the bulk.

\item \textbf{$\mu$ and B$\mu$-term}: The $\mu$-term can be generated by a standard Giudice-Masiero term in the K\"ahler potential
$K \supset \tilde{K}(t_s,t_b) H_u H_d$ which gives again $\mu\simeq m_{3/2}$ and $B\mu \simeq m_{3/2}^2$.
\end{itemize}
Summarising, we obtained a very specific pattern of soft terms with scalars heavier than the gauginos and
universal A-terms and $\mu$/$B\mu$-term of the order the gravitino mass:
\be
m_0 \simeq A_{\alpha\beta\gamma}\simeq \mu\simeq B \simeq m_{3/2} \gg M_{1/2} \simeq \delta \,m_{3/2}\,.
\ee
The soft masses scale with $m_{3/2}$ and do not depend on the mechanism which stabilises the complex structure and bundle moduli.
Hence one can obtain TeV-scale supersymmetry by considering either smooth CY models where all
the complex structure moduli are fixed by the holomorphicity of the gauge bundle or orbifold constructions
with a small number of untwisted $Z$-moduli (or better with no untwisted $Z$-moduli at all as in the case
of some non-Abelian orbifolds). On general Calabi-Yau manifolds, we expect the soft mass scale to be of order $m_{3/2}\sim M_{\rm \GUT}$ due the fact that in the heterotic string there is not enough freedom to tune the flux superpotential below values of $\mc{O}(0.1 - 0.01)$.

\section{Anisotropic solutions}
\label{sec:Anisotropic}

In this section we shall show how to generalise the previous results to obtain anisotropic compactifications
with 2 large and 4 small extra dimensions which allow for a right value of the GUT scale.\footnote{For anisotropic solutions in the type IIB case for the same
kind of fibred CY manifolds see \cite{AnisotropicLVS1,AnisotropicLVS2,AnisotropicLVS3}.} For this purpose,
we shall focus on CY three-folds whose volume is \cite{CKM}:
\be
\vo = k_b t_b t_f^2 - k_s t_s^3\,.
\label{FibredVol}
\ee
This CY admits a 4D K3 or $T^4$ divisor of volume $t_f^2$ fibered over a 2D $\mathbb{P}^1$ base of volume $t_b$
with an additional del Pezzo divisor of size $t_s^2$. We shall now show how to fix the moduli dynamically
in the anisotropic region $t_b \gg t_f \sim t_s$. We shall consider a hidden sector gauge kinetic function of the form:
\be
f_{\rm hid} = S - \frac{\beta_b}{2}\,T_b - \frac{\beta_f}{2}\,T_f \equiv \Phi \,,\qquad\text{with}\qquad\beta_s=0\,.
\ee
The superpotential looks exactly as the one in (\ref{Wsimple}) whereas the K\"ahler potential reads:
\be
K= -\ln\vo-\ln\left[\Phi+\bar{\Phi}+\frac{\beta_b}{2}\left(T_b+\bar{T}_b\right)+\frac{\beta_f}{2}\left(T_f+\bar{T}_f\right)\right]+K_{\rm cs}(Z)\,.
\ee
Focusing on the limit where 1-loop effects are suppressed with respect to the tree-level expression of
the gauge kinetic function:
\be
\epsilon_b \equiv \frac{\beta_b\,t_b}{2\phi} \ll 1\qquad\text{and}\qquad
\epsilon_f \equiv \frac{\beta_f\,t_f}{2\phi} \ll 1\,,
\ee
the dilaton is again fixed at leading order by requiring $D_\Phi W=0$.
On the other hand the K\"ahler moduli develop a subdominant potential
via non-perturbative contributions, $\alpha'$ corrections and threshold
effects which break the no-scale structure as:
\be
\sum_T K^{i\bar{j}} K_i  K_{\bar{j}} -3 = 2\left(\epsilon_b+\epsilon_f\right) + \mc{O}(\epsilon^2)\,.
\ee
The scalar potential has therefore the same expression as (\ref{VOe}) but with a different
coefficient $A_1$ which is now moduli-dependent and looks like:
\be
A_1 (\vo,t_f) = \frac{\vo^{2/3}}{2\langle\phi\rangle}\left(\frac{\beta_b}{k_b t_f^2}+\frac{\beta_f\,t_f}{\vo}\right)\,,
\label{A1new}
\ee
where we have traded $t_b$ for $\vo$. This is the only term which depends on $t_f$ since the rest of the potential depends
just on $\vo$ and $t_s$. Hence we can fix $t_f$ just minimising $A_1(\vo,t_f)$ obtaining:
\be
t_f = \left(\frac{2\beta_b}{k_b \beta_f}\right)^{1/3} \vo^{1/3}\qquad\Leftrightarrow\qquad
t_f = \frac{2\beta_b}{\beta_f} \,t_b\,. \label{Anis}
\ee
Substituting this result in (\ref{A1new}) we find that $A_1$ becomes:
\be
A_1 = \frac{3 \beta_b}{2\langle\phi\rangle k_b^{1/3}}\left(\frac{\beta_f}{2\beta_b}\right)^{2/3}\,,
\ee
which is not moduli-dependent anymore and takes a form very similar to the one in (\ref{Param}).
We can therefore follow the same stabilisation procedure described in the previous sections but now with
the additional relation (\ref{Anis}) which, allowing the moderate tuning $\beta_f \simeq 20\,\beta_b$,
would give an anisotropic solution with $t_b \simeq 10 \,t_f$. For example for $\vo \simeq 20$ and $k_b=1/2$,
one would obtain $t_b\simeq 16 \gg t_f \simeq 1.6$.

We finally mention that this kind of fibred CY manifolds have been successfully used in type IIB for deriving inflationary models
from string theory where the inflaton is the K\"ahler modulus controlling the volume of the fibre \cite{FibreInflation}.
It would be very interesting to investigate if similar cosmological applications could also be present in the heterotic case.

\section{Conclusions}
\label{Conclusions}

The heterotic string on a CY manifold (or its various limiting
cases such as orbifolds and Gepner points) has been studied since
the late eighties as a possible UV complete theory of gravity that can
realise a unified version of the SM. In the last decade
there has been much progress towards the goal of
getting a realistic model with the correct spectrum. However the major problem in getting phenomenologically viable solutions for
the heterotic string is that the gauge theory
resides in the bulk, and so getting an acceptable model cannot be decoupled
from the problem of moduli stabilisation.
Unfortunately a complete and deep understanding of the mechanism
which stabilises all the moduli in the heterotic string is still lacking.

In this paper we tried to perform a systematic analysis of all the effects which
can develop a potential for the various moduli for the case of $(0,2)$-compactifications
which allow for MSSM-like model building and the generation of worldsheet instantons
that are crucial effects to fix the K\"ahler moduli. According to the original Strominger's analysis \cite{Strominger:1986uh},
these compactifications violate the K\"ahler condition $dJ=0$
due to a non-zero $H$-flux at $\mc{O}(\alpha')$ since in the non-standard embedding
the co-exact piece of the Chern-Simons term in $H$ does not cancel.
We then considered solutions to the 10D equations of motion with constant dilaton and warp factor,
corresponding to `special Hermitian manifolds', which represent the smallest deviations
from smooth CY manifolds at $\mc{O}(\alpha')$ \cite{Lopes Cardoso:2002hd}.

Let us summarise the various moduli stabilisation effects that we have taken into account:
\begin{itemize}
\item \textit{Holomorphicity of the gauge bundle, D-terms and higher order perturbative contributions to the superpotential}:
By demanding a supersymmetric gauge bundle, i.e. a gauge bundle which satisfies the Hermitian Yang-Mills equations,
the combined space of complex structure and gauge bundle moduli reduces from a naive direct product
to a `cross-structure' \cite{AGLO1,AGLO2,Anderson:2011ty,Witten:1985bz}.
Therefore if the gauge bundle moduli are fixed at non-zero VEVs by D-terms combined with higher order
perturbative contributions to the superpotential \cite{Buchmuller:2005jr,Buchmuller:2006ik,Lebedev:2006kn,Lebedev:2007hv,Lebedev:2008un},
the $Z$-moduli are automatically lifted.
However, not all the complex structure moduli might get frozen by this mechanism
since, in general, the sub-locus in complex structure moduli space where the gauge bundle is holomorphic
turns out to have dimension $h^{1,2}_{\rm hol}>0$. Hence $0<h^{1,2}_{\rm hol}<h^{1,2}$ flat $Z$-moduli are generically left over.

\item \textit{Fractional Chern-Simons invariants, gaugino condensation and threshold effects}:
The remaining flat $Z$-directions could be lifted by turning on quantised background three-form fluxes \cite{GKLM,BdA,CKL}.
However we showed that, contrary to type IIB, this cannot be done having at the same time a vanishing VEV
of the tree-level flux superpotential $W_0$ since setting the F-terms of the $Z$-moduli to zero corresponds
to setting the $(1,2)$-component of the $H$-flux to zero, while demanding $W_0=0$ implies that also the $(3,0)$-piece
of $H$ is vanishing. Hence, being real, the whole $H$-flux has to be zero, resulting in the impossibility of fixing
the remaining complex structure moduli. Thus one needs $W_0\neq 0$ in order to fix the $Z$-moduli.
However, due to the absence of Ramond-Ramond fluxes, it is hard to tune $W_0$ small enough
to balance the exponentially suppressed contribution from gaugino condensation which introduces an
explicit dependence on the dilaton~\cite{Font:1990nt,Ferrara:1990ei,Nilles:1990jv}
unless one turns on fractional Chern-Simons invariants
(i.e. discrete Wilson lines) \cite{GKLM}. In this way both the dilaton and the complex structure moduli can be stabilised
supersymmetrically at non-perturbative level. The K\"ahler moduli could then be fixed
by the inclusion of threshold corrections to the gauge kinetic function \cite{Dixon:1990pc,BHW}.

\item \textit{Worldsheet instantons}:
The supersymmetric minimum obtained by including threshold effects is not in the weak coupling regime where one can trust the effective field
theory. This problem can be avoided by considering also the contribution of $T$-dependent worldsheet instantons
which can give rise to reliable supersymmetric AdS vacua \cite{CKL}.

\item \textit{Higher derivative and loop corrections to the K\"ahler potential}:
The last effects to be taken into account are $\alpha'$ corrections to the K\"ahler potential \cite{AQS,StandardAlphaPrime,BBHL},
while string loop effects can be estimated to give rise to negligible contributions to the scalar potential \cite{BHK,BHP,CCQ}.
These higher derivative corrections yield new stable vacua where supersymmetry is spontaneously broken
by the stabilisation mechanism which induces non-zero F-terms for the K\"ahler moduli, in a way very similar
to type IIB LARGE Volume Scenarios \cite{LVS,GeneralLVS}. These new vacua can be Minkowski due to the positive
contribution from threshold effects. However, due to the lack of tuning freedom in $W_0$,
it is very hard to achieve vacua with exponentially small vacuum energy. Thus we used the term `Minkowski vacua'
to refer to solutions with a cosmological constant suppressed by at least a loop factor with respect to
the height of the barrier in the scalar potential which prevents the K\"ahler moduli to run-away to infinity.
Moreover, this stabilisation mechanism allows for anisotropic compactifications
with two extra dimensions which are much larger than the other four. In this way,
the unification scale can be lowered down to the observed phenomenological value~\cite{Hebecker:2004ce,Dundee:2008ts},
fitting very well with the picture of 6D orbifold GUTs~\cite{Dundee:2008ts,Buchmuller:2007qf}.
\end{itemize}

After showing the existence of this new kind of supersymmetry breaking vacua, we estimated
the size of the soft terms generated by gravity mediation. Interestingly,
they feature universal scalar masses, A-terms and $\mu/B\mu$-term of
$\mc{O}(m_{3/2})$ and suppressed gaugino masses at the \%-level.
Moreover, a potentially viable QCD axion candidate is given by the axionic partner of the `large' 2-cycle modulus.
However, due to the lack of tuning freedom in the flux superpotential $W_0\simeq \mc{O}(0.1-0.01)$,
the gravitino mass $m_{3/2} = W_0 M_P/\sqrt{2\rm{Re}(S)\vo}$ becomes of order $M_{\GUT}\simeq 10^{16}$ GeV for $\rm{Re}(S)\simeq 2$
and $\vo\simeq 20$. This is not a problem if one does not believe in the solution of the hierarchy problem based
on low-energy supersymmetry, but it represents a generic prediction of weakly coupled heterotic compactifications
on internal manifolds which are smooth CY three-folds up to $\alpha'$ effects.

However, our stabilisation procedure for the K\"ahler moduli that leads to spontaneous supersymmetry breaking,
is completely independent on the supersymmetric mechanism which is used to fix the dilaton and the complex structure moduli.
Hence, if one is instead interested in low-energy supersymmetry, our way to break supersymmetry along the K\"ahler moduli
directions could still be used by focusing on different ways to freeze the $S$- and $Z$-moduli:
\begin{enumerate}
\item In some particular examples all the complex structure moduli could be stabilised by
the requirement of a holomorphic gauge bundle \cite{AGLO1,AGLO2,Anderson:2011ty}. In this case one could have $W_0=0$ and an exponentially
small superpotential, leading to a TeV-scale gravitino mass, could be generated by gaugino condensation.

\item In Abelian orbifold models the number of untwisted complex structure moduli is very small.
There are also some non-Abelian orbifolds with no $Z$-moduli at all. Hence in this case
it is rather likely that all the $Z$-moduli could be fixed by the holomorphicity of the gauge bundle
once all the singlets are fixed at non-zero VEVs by cancelling FI-terms or by the effect of
higher order terms in $W$ \cite{Buchmuller:2005jr,Buchmuller:2006ik,Lebedev:2006kn,Lebedev:2007hv,Lebedev:2008un}.
Again, $m_{3/2}$ could then be lowered to the TeV-scale due to the exponential suppression coming from
gaugino condensation \cite{HPN,Kappl:2008ie}.

\item The flux superpotential could have enough tuning freedom in the presence of fluxes
which are the equivalent of type IIB Ramond-Ramond fluxes. This is the case
of non-complex manifolds with new geometric fluxes where the $H$ flux gets modified to
$\mc{H} = H+{\rm i}\,dJ$ \cite{Lopes Cardoso:2002hd,CCDL, Held:2010az, Becker:2003gq, Becker:2003yv}.\footnote{See also \cite{Becker:2009df}.}
\end{enumerate}
We finally stress that, even if these models could give low-energy supersymmetry,
the possibility to tune the cosmological constant to the observed value still remains a challenge,
in particular in the cases without a large flux discretuum.

\section*{Acknowledgments}

We would like to thank Lara Anderson, Lilia Anguelova, Arthur Hebecker, James Gray, Shamit Kachru, Oleg Lebedev, Andre Lukas, Anshuman Maharana, Hans Peter Nilles, Francisco Pedro, Fernando Quevedo, Callum Quigley, Stuart Raby, Marco Serone, Savdeep Sethi, Patrick Vaudrevange, Roberto Zucchini and especially Roberto Valandro for useful discussions and correspondence. We are grateful for the support of, and the pleasant environs provided by, the Abdus Salam International Center for Theoretical Physics. This work was supported by the Impuls und Vernetzungsfond of the Helmholtz Association of German Research Centres under grant HZ-NG-603, and German Science Foundation (DFG) within the Collaborative Research Center (CRC) 676 ÓParticles, Strings and the Early UniverseÓ, and by the United States Department of Energy under grant DE-FG02-91-ER-40672. SdA would like to acknowledge the award of an CRC 676 fellowship from DESY/Hamburg University and a visiting professorship at Abdus Salam ICTP.

\begin{appendix}

\section{Dimensional reduction of 10D heterotic action}
\label{AppA}

The 10D heterotic supergravity action in string frame for energies below the mass of the first excited string state
$M_s= \ell_s^{-1}$ with $\ell_s=2\pi\sqrt{\alpha'}$ contains bosonic terms of the form:
\begin{eqnarray}
S&\supset&\frac{1}{(2\pi)^7 \alpha'^4}\int d^{10}x \sqrt{-G} \,e^{-2\phi} \left(\mc{R}-\frac{\alpha'}{4}{\rm Tr} F^2\right) \nn \\
&=& \frac{M_\10^8}{2}\int d^{10}x \sqrt{-G} \,e^{-2\phi}\mc{R} -\frac{1}{2 g_\10^2}\int d^{10}x \sqrt{-G}\, e^{-2\phi} {\rm Tr} F^2\,.
\label{10Daction}
\end{eqnarray}
Comparing the first with the second line in (\ref{10Daction}), we find:
\be
M_\10^8 = \frac{2}{(2\pi)^7 \alpha'^4} = 4\pi M_s^8 \qquad\text{and}\qquad g_\10^2 = \frac{4}{\alpha' M_\10^8} = 4 \pi M_s^{-6}\,.
\ee
Compactifying on a 6D CY three-fold $X$, the 4D Planck scale $M_P$ turns out to be:
\be
M_P^2 = e^{-2\langle\phi\rangle} M_\10^8 {\rm Vol}(X)= 4\pi\,g_s^{-2}\vo M_s^2\,,
\label{MP}
\ee
where we measured the internal volume in units of $M_s^{-1}$ as ${\rm Vol}(X)=\vo \,\ell_s^6$ and
we explicitly included factors of the string coupling $g_s=e^{\langle\phi\rangle}$.
On the other hand, the 4D gauge coupling constant becomes:
\be
\alpha_\GUT^{-1} = 4\pi g_\4^{-2} = \frac{4\pi {\rm Vol}(X)}{g_\10^2 \,e^{2\langle\phi\rangle}}=
g_s^{-2}\vo\,.
\label{alphaGUT}
\ee
The tree-level expression of the gauge kinetic function $f=S$ requires ${\rm Re}(S)=g_4^{-2}$, implying the following
normalisation of the definition of the dilaton field:
\be
S= \frac{1}{4\pi} \left(e^{-2\phi}\vo+{\rm i} \,a\right).
\ee
From (\ref{alphaGUT}), we immediately realise that there is a tension between large volume and weak coupling
for the physical value $\alpha_\GUT^{-1}\simeq 25$:
\be
\vo = g_s^2 \alpha_\GUT^{-1} \simeq g_s^2 25 \lesssim 25 \qquad\text{for}\qquad g_s\lesssim 1\,.
\ee
On top of this problem, isotropic compactifications cannot yield the right value of the GUT scale $M_\GUT\simeq 2.1 \cdot 10^{16}$ GeV
which is given by the Kaluza-Klein scale $M_{\KK} = M_s/\vo^{1/6}$. In fact, combining (\ref{MP}) with (\ref{alphaGUT}),
one finds that the string scale is fixed to be very high:
\be
M_s^2 = \frac{M_P^2}{4\pi\alpha_\GUT^{-1}} \simeq \frac{M_P^2}{100\pi}\simeq \left(1.35\cdot 10^{17}\,\text{GeV}\right)^2\,.
\ee
In turn, for $\vo\lesssim 25$, the GUT scale becomes too high: $M_\GUT=M_\KK \gtrsim 8\cdot 10^{16}$ GeV.
The situation can be improved by focusing on anisotropic compactifications with $d$ large extra dimensions of size $L=x\ell_s$ with $x\gg 1$
and $(6-d)$ small dimensions of string size $l=\ell_s$. The internal volume then becomes ${\rm Vol}(X)= L^d l^{(6-d)} = x^d \ell_s^6=\vo \,\ell_s^6$,
implying that the Kaluza-Klein scale now becomes $M_\KK = M_s/x = M_s/\vo^{1/d}$. Clearly, for the case $d=6$, we recover the isotropic situation.
The case with $d=1$ is not very interesting since CY manifolds do not admit non-trivial Wilson lines to perform the GUT breaking.
We shall therefore focus on the case $d=2$ where we get the promising result:
\be
M_\GUT=M_\KK = \frac{M_s}{\sqrt{\vo}} \gtrsim 2.7\cdot 10^{16}\,\text{GeV}.
\ee

\end{appendix}


\begin{thebibliography}{99}
\bibitem{Grana:2005jc}
  M.~Grana,
  ``Flux compactifications in string theory: A Comprehensive review,''
  Phys.\ Rept.\  {\bf 423} (2006) 91
  [hep-th/0509003].

\bibitem{Douglas:2006es}
  M.~R.~Douglas and S.~Kachru,
  ``Flux compactification,''
  Rev.\ Mod.\ Phys.\  {\bf 79} (2007) 733
  [hep-th/0610102].

\bibitem{Braun:2005ux}
  V.~Braun, Y.~-H.~He, B.~A.~Ovrut and T.~Pantev,
  ``A Heterotic standard model,''
  Phys.\ Lett.\ B {\bf 618}, 252 (2005)
  [hep-th/0501070].

\bibitem{Buchmuller:2005jr}
  W.~Buchmuller, K.~Hamaguchi, O.~Lebedev and M.~Ratz,
  ``Supersymmetric standard model from the heterotic string,''
  Phys.\ Rev.\ Lett.\  {\bf 96}, 121602 (2006)
  [hep-ph/0511035].

\bibitem{Buchmuller:2006ik}
  W.~Buchmuller, K.~Hamaguchi, O.~Lebedev and M.~Ratz,
  ``Supersymmetric Standard Model from the Heterotic String (II),''
  Nucl.\ Phys.\ B {\bf 785}, 149 (2007)
  [hep-th/0606187].

\bibitem{Lebedev:2006kn}
  O.~Lebedev, H.~P.~Nilles, S.~Raby, S.~Ramos-Sanchez, M.~Ratz, P.~K.~S.~Vaudrevange and A.~Wingerter,
  ``A Mini-landscape of exact MSSM spectra in heterotic orbifolds,''
  Phys.\ Lett.\ B {\bf 645}, 88 (2007)
  [hep-th/0611095].

\bibitem{Lebedev:2007hv}
  O.~Lebedev, H.~P.~Nilles, S.~Raby, S.~Ramos-Sanchez, M.~Ratz, P.~K.~S.~Vaudrevange and A.~Wingerter,
  ``The Heterotic Road to the MSSM with R parity,''
  Phys.\ Rev.\ D {\bf 77}, 046013 (2008)
  [arXiv:0708.2691 [hep-th]].

\bibitem{Lebedev:2008un}
  O.~Lebedev, H.~P.~Nilles, S.~Ramos-Sanchez, M.~Ratz and P.~K.~S.~Vaudrevange,
  ``Heterotic mini-landscape. (II). Completing the search for MSSM vacua in a Z(6) orbifold,''
  Phys.\ Lett.\ B {\bf 668}, 331 (2008)
  [arXiv:0807.4384 [hep-th]].

\bibitem{Dundee:2010sb}
  B.~Dundee, S.~Raby and A.~Westphal,
  ``Moduli stabilization and SUSY breaking in heterotic orbifold string models,''
  Phys.\ Rev.\ D {\bf 82}, 126002 (2010)
  [arXiv:1002.1081 [hep-th]].

\bibitem{Parameswaran:2010ec}
  S.~L.~Parameswaran, S.~Ramos-Sanchez and I.~Zavala,
  ``On Moduli Stabilisation and de Sitter Vacua in MSSM Heterotic Orbifolds,''
  JHEP {\bf 1101}, 071 (2011)
  [arXiv:1009.3931 [hep-th]].

\bibitem{Font:1990nt}
  A.~Font, L.~E.~Ibanez, D.~Lust and F.~Quevedo,
  ``Supersymmetry Breaking From Duality Invariant Gaugino Condensation,''
  Phys.\ Lett.\ B {\bf 245}, 401 (1990).

\bibitem{Ferrara:1990ei}
  S.~Ferrara, N.~Magnoli, T.~R.~Taylor and G.~Veneziano,
  ``Duality and supersymmetry breaking in string theory,''
  Phys.\ Lett.\ B {\bf 245}, 409 (1990).

\bibitem{Nilles:1990jv}
  H.~P.~Nilles and M.~Olechowski,
  ``Gaugino Condensation And Duality Invariance,''
  Phys.\ Lett.\ B {\bf 248}, 268 (1990).

\bibitem{Casas:1990qi}
  J.~A.~Casas, Z.~Lalak, C.~Munoz and G.~G.~Ross,
  ``Hierarchical Supersymmetry Breaking And Dynamical Determination Of Compactification Parameters By Nonperturbative Effects,''
  Nucl.\ Phys.\ B {\bf 347}, 243 (1990).

\bibitem{de Carlos:1992da}
  B.~de Carlos, J.~A.~Casas and C.~Munoz,
  ``Supersymmetry breaking and determination of the unification gauge coupling constant in string theories,''
  Nucl.\ Phys.\ B {\bf 399}, 623 (1993)
  [hep-th/9204012].

\bibitem{Gallego:2008sv}
  D.~Gallego and M.~Serone,
  ``Moduli Stabilization in non-Supersymmetric Minkowski Vacua with Anomalous U(1) Symmetry,''
  JHEP {\bf 0808}, 025 (2008)
  [arXiv:0807.0190 [hep-th]].

\bibitem{GKLM}
  S.~Gukov, S.~Kachru, X.~Liu and L.~McAllister,
  ``Heterotic moduli stabilization with fractional Chern-Simons invariants,''
  Phys.\ Rev.\ D {\bf 69} (2004) 086008  [hep-th/0310159].  

\bibitem{BdA}
  R.~Brustein and S.~P.~de Alwis,
  ``Moduli potentials in string compactifications with fluxes: Mapping the discretuum,''
  Phys.\ Rev.\ D {\bf 69} (2004) 126006  [hep-th/0402088].  

\bibitem{CKL}
  G.~Curio, A.~Krause and D.~Lust,
  ``Moduli stabilization in the heterotic/IIB discretuum,''
  Fortsch.\ Phys.\  {\bf 54} (2006) 225  [hep-th/0502168].  

\bibitem{Strominger:1986uh}
  A.~Strominger,
  ``Superstrings with Torsion,''
  Nucl.\ Phys.\ B {\bf 274} (1986) 253.

\bibitem{Dixon:1990pc}
  L.~J.~Dixon, V.~Kaplunovsky and J.~Louis,
  ``Moduli dependence of string loop corrections to gauge coupling constants,''
  Nucl.\ Phys.\ B {\bf 355}, 649 (1991).

\bibitem{BHW}
  R.~Blumenhagen, G.~Honecker and T.~Weigand,
  ``Loop-corrected compactifications of the heterotic string with line bundles,''
  JHEP {\bf 0506} (2005) 020  [hep-th/0504232].  

\bibitem{AQS}
  L.~Anguelova, C.~Quigley and S.~Sethi,
  ``The Leading Quantum Corrections to Stringy Kahler Potentials,''
  JHEP {\bf 1010} (2010) 065  [arXiv:1007.4793 [hep-th]].  

\bibitem{StandardAlphaPrime}
  P.~Candelas, X.~C.~De La Ossa, P.~S.~Green and L.~Parkes,
  ``A Pair of Calabi-Yau manifolds as an exactly soluble superconformal theory,''
  Nucl.\ Phys.\ B {\bf 359} (1991) 21.  

\bibitem{BBHL}
  K.~Becker, M.~Becker, M.~Haack and J.~Louis,
  ``Supersymmetry breaking and alpha-prime corrections to flux induced potentials,''
  JHEP {\bf 0206} (2002) 060  [hep-th/0204254].  

\bibitem{LVS}
  V.~Balasubramanian, P.~Berglund, J.~P.~Conlon and F.~Quevedo,
  ``Systematics of moduli stabilisation in Calabi-Yau flux compactifications,''
  JHEP {\bf 0503} (2005) 007  [hep-th/0502058].  

\bibitem{GeneralLVS}
  M.~Cicoli, J.~P.~Conlon and F.~Quevedo,
  ``General Analysis of LARGE Volume Scenarios with String Loop Moduli Stabilisation,''
  JHEP {\bf 0810} (2008) 105  [arXiv:0805.1029 [hep-th]].  

\bibitem{Hebecker:2004ce}
  A.~Hebecker and M.~Trapletti,
  ``Gauge unification in highly anisotropic string compactifications,''
  Nucl.\ Phys.\ B {\bf 713}, 173 (2005)
  [hep-th/0411131].

\bibitem{Dundee:2008ts}
  B.~Dundee, S.~Raby and A.~Wingerter,
  ``Reconciling Grand Unification with Strings by Anisotropic Compactifications,''
  Phys.\ Rev.\ D {\bf 78}, 066006 (2008)
  [arXiv:0805.4186 [hep-th]].

\bibitem{Buchmuller:2007qf}
  W.~Buchmuller, C.~Ludeling and J.~Schmidt,
  ``Local SU(5) Unification from the Heterotic String,''
  JHEP {\bf 0709}, 113 (2007)
  [arXiv:0707.1651 [hep-ph]].

\bibitem{HPN}
  O.~Loaiza-Brito, J.~Martin, H.~P.~Nilles and M.~Ratz,
  ``Log(M(Pl) / m(3/2)),''
  AIP Conf.\ Proc.\  {\bf 805} (2006) 198
  [hep-th/0509158].

\bibitem{Kappl:2008ie}
  R.~Kappl, H.~P.~Nilles, S.~Ramos-Sanchez, M.~Ratz, K.~Schmidt-Hoberg and P.~K.~S.~Vaudrevange,
  ``Large hierarchies from approximate R symmetries,''
  Phys.\ Rev.\ Lett.\  {\bf 102}, 121602 (2009)
  [arXiv:0812.2120 [hep-th]].

\bibitem{AGLO1}
  L.~B.~Anderson, J.~Gray, A.~Lukas and B.~Ovrut,
  ``Stabilizing the Complex Structure in Heterotic Calabi-Yau Vacua,''
  JHEP {\bf 1102} (2011) 088  [arXiv:1010.0255 [hep-th]].  

\bibitem{AGLO2}
  L.~B.~Anderson, J.~Gray, A.~Lukas and B.~Ovrut,
  ``Stabilizing All Geometric Moduli in Heterotic Calabi-Yau Vacua,''
  Phys.\ Rev.\ D {\bf 83} (2011) 106011  [arXiv:1102.0011 [hep-th]].  

\bibitem{Anderson:2011ty}
  L.~B.~Anderson, J.~Gray, A.~Lukas and B.~Ovrut,
  ``The Atiyah Class and Complex Structure Stabilization in Heterotic Calabi-Yau Compactifications,''
  JHEP {\bf 1110} (2011) 032 [arXiv:1107.5076 [hep-th]].

\bibitem{Witten:1985bz}
  E.~Witten,
  ``New Issues in Manifolds of SU(3) Holonomy,''
  Nucl.\ Phys.\ B {\bf 268} (1986) 79.

\bibitem{Donagi}
  R.~Donagi and M.~Wijnholt,
  ``Higgs Bundles and UV Completion in F-Theory,''
  arXiv:0904.1218 [hep-th];
R.~Donagi and M.~Wijnholt,
  ``Model Building with F-Theory,''
  Adv.\ Theor.\ Math.\ Phys.\  {\bf 15}, 1237 (2011)
  [arXiv:0802.2969 [hep-th]].

\bibitem{AGLO3}
  L.~B.~Anderson, J.~Gray, A.~Lukas and B.~Ovrut,
  ``Vacuum Varieties, Holomorphic Bundles and Complex Structure Stabilization in Heterotic Theories,''
  JHEP {\bf 1307} (2013) 017  [arXiv:1304.2704 [hep-th]].

\bibitem{Lopes Cardoso:2002hd}
  G.~Lopes Cardoso, G.~Curio, G.~Dall'Agata, D.~Lust, P.~Manousselis and G.~Zoupanos,
  ``NonKahler string backgrounds and their five torsion classes,''
  Nucl.\ Phys.\ B {\bf 652}, 5 (2003)
  [hep-th/0211118].

\bibitem{GKP}
  S.~B.~Giddings, S.~Kachru and J.~Polchinski,
  ``Hierarchies from fluxes in string compactifications,''
  Phys.\ Rev.\ D {\bf 66} (2002) 106006  [hep-th/0105097].  

\bibitem{CCDL}
  G.~Lopes Cardoso, G.~Curio, G.~Dall'Agata and D.~Lust,
  ``BPS action and superpotential for heterotic string compactifications with fluxes,''
  JHEP {\bf 0310} (2003) 004  [hep-th/0306088].  

\bibitem{Held:2010az}
  J.~Held, D.~Lust, F.~Marchesano and L.~Martucci,
  ``DWSB in heterotic flux compactifications,''
  JHEP {\bf 1006} (2010) 090
  [arXiv:1004.0867 [hep-th]].

\bibitem{Becker:2003gq}
  K.~Becker, M.~Becker, K.~Dasgupta and S.~Prokushkin,
  ``Properties of heterotic vacua from superpotentials,''
  Nucl.\ Phys.\ B {\bf 666} (2003) 144
  [hep-th/0304001].

\bibitem{Becker:2003yv}
  K.~Becker, M.~Becker, K.~Dasgupta and P.~S.~Green,
  ``Compactifications of heterotic theory on nonKahler complex manifolds. 1.,''
  JHEP {\bf 0304} (2003) 007
  [hep-th/0301161].

\bibitem{Witten:1985xb}
  E.~Witten,
  ``Dimensional Reduction of Superstring Models,''
  Phys.\ Lett.\ B {\bf 155}, 151 (1985).

\bibitem{Ferrara:1986qn}
  S.~Ferrara, C.~Kounnas and M.~Porrati,
  ``General Dimensional Reduction of Ten-Dimensional Supergravity and Superstring,''
  Phys.\ Lett.\ B {\bf 181}, 263 (1986).

\bibitem{BHK}
  M.~Berg, M.~Haack and B.~Kors,
  ``String loop corrections to Kahler potentials in orientifolds,''
  JHEP {\bf 0511} (2005) 030  [hep-th/0508043].  

\bibitem{BHP}
  M.~Berg, M.~Haack and E.~Pajer,
  ``Jumping Through Loops: On Soft Terms from Large Volume Compactifications,''
  JHEP {\bf 0709} (2007) 031  [arXiv:0704.0737 [hep-th]].  

\bibitem{CCQ}
  M.~Cicoli, J.~P.~Conlon and F.~Quevedo,
  ``Systematics of String Loop Corrections in Type IIB Calabi-Yau Flux Compactifications,''
  JHEP {\bf 0801} (2008) 052  [arXiv:0708.1873 [hep-th]].  

\bibitem{Candelas:1990pi}
  P.~Candelas and X.~de la Ossa,
  ``Moduli Space Of Calabi-yau Manifolds,''
  Nucl.\ Phys.\ B {\bf 355} (1991) 455.

\bibitem{Greene:1990ud}
  B.~R.~Greene and M.~R.~Plesser,
  ``Duality In Calabi-yau Moduli Space,''
  Nucl.\ Phys.\ B {\bf 338}, 15 (1990).

\bibitem{Candelas:1994hw}
  P.~Candelas, A.~Font, S.~H.~Katz and D.~R.~Morrison,
  ``Mirror symmetry for two parameter models. 2.,''
  Nucl.\ Phys.\ B {\bf 429}, 626 (1994)
  [hep-th/9403187].

\bibitem{Candelas:2000fq}
  P.~Candelas, X.~de la Ossa and F.~Rodriguez-Villegas,
  ``Calabi-Yau manifolds over finite fields. 1.,''
  hep-th/0012233.

\bibitem{Giryavets:2003vd}
  A.~Giryavets, S.~Kachru, P.~K.~Tripathy and S.~P.~Trivedi,
  ``Flux compactifications on Calabi-Yau threefolds,''
  JHEP {\bf 0404}, 003 (2004)
  [hep-th/0312104].

\bibitem{Denef:2004dm}
  F.~Denef, M.~R.~Douglas and B.~Florea,
  ``Building a better racetrack,''
  JHEP {\bf 0406}, 034 (2004)
  [hep-th/0404257].

\bibitem{AQ}
  L.~Anguelova and C.~Quigley,
  ``Quantum Corrections to Heterotic Moduli Potentials,''
  JHEP {\bf 1102} (2011) 113  [arXiv:1007.5047 [hep-th]].  

\bibitem{Kmatter}
  J.~P.~Conlon, D.~Cremades and F.~Quevedo,
  ``Kahler potentials of chiral matter fields for Calabi-Yau string compactifications,''
  JHEP {\bf 0701} (2007) 022  [hep-th/0609180].  

\bibitem{IIBSoftTerms}
  J.~P.~Conlon, S.~S.~Abdussalam, F.~Quevedo and K.~Suruliz,
  ``Soft SUSY Breaking Terms for Chiral Matter in IIB String Compactifications,''
  JHEP {\bf 0701} (2007) 032  [hep-th/0610129].  

\bibitem{AnisotropicLVS1}
  M.~Cicoli, C.~P.~Burgess and F.~Quevedo,
  ``Anisotropic Modulus Stabilisation: Strings at LHC Scales with Micron-sized Extra Dimensions,''
  JHEP {\bf 1110} (2011) 119  [arXiv:1105.2107 [hep-th]].  

\bibitem{AnisotropicLVS2}
  M.~Cicoli, C.~Mayrhofer and R.~Valandro,
  ``Moduli Stabilisation for Chiral Global Models,''
  JHEP {\bf 1202} (2012) 062  [arXiv:1110.3333 [hep-th]].  

\bibitem{AnisotropicLVS3}
  S.~Angus and J.~P.~Conlon,
  ``Soft Supersymmetry Breaking in Anisotropic LARGE Volume Compactifications,''
  arXiv:1211.6927 [hep-th].  

\bibitem{CKM}
  M.~Cicoli, M.~Kreuzer and C.~Mayrhofer,
  ``Toric K3-Fibred Calabi-Yau Manifolds with del Pezzo Divisors for String Compactifications,''
  JHEP {\bf 1202} (2012) 002  [arXiv:1107.0383 [hep-th]].  

\bibitem{FibreInflation}
  M.~Cicoli, C.~P.~Burgess and F.~Quevedo,
  ``Fibre Inflation: Observable Gravity Waves from IIB String Compactifications,''
  JCAP {\bf 0903} (2009) 013  [arXiv:0808.0691 [hep-th]].  

\bibitem{Becker:2009df}
  K.~Becker and S.~Sethi,
  ``Torsional Heterotic Geometries,''
  Nucl.\ Phys.\ B {\bf 820} (2009) 1
  [arXiv:0903.3769 [hep-th]].

\end{thebibliography}
\end{document}